\documentclass[11pt]{article}
\usepackage{amsmath,amsthm,latexsym,amssymb,amsfonts,epsfig,psfrag,color}
\usepackage{subfigure}
\usepackage{graphicx,overpic}

\usepackage{colortbl}

\usepackage[sort&compress,numbers]{natbib}

\usepackage{booktabs}

\definecolor{linkc}{rgb}{0,0,0.3}
\definecolor{rule}{rgb}{0.7,0.7,0.7}

\usepackage{hyperref}
 \hypersetup{colorlinks=true, linkcolor=black, urlcolor=linkc, citecolor=linkc, bookmarksopen}

\makeatletter \@addtoreset{equation}{section} \makeatother

\def\be{\begin{equation}}
\def\ee{\end{equation}}
\def\ba{\begin{eqnarray}}
\def\ea{\end{eqnarray}}

\newcommand\nn{\nonumber}
\newcommand\q{\quad}
\newcommand{\Zl}{\mathbb{Z}}

\newcommand{\varep}{\varepsilon}

\renewenvironment{thebibliography}[1]{%
\begin{oldthebibliography}{#1}%
\setlength{\parskip}{0pt}%
\setlength{\itemsep}{0pt}%
}%
{%
\end{oldthebibliography}%
}

\title{Coarse graining methods for spin net and spin foam models}
\author{
 Bianca Dittrich, Frank C.~Eckert, Mercedes Martin-Benito\\
\small MPI f. Gravitational Physics, Albert Einstein Institute,\\
 \small Am M\"uhlenberg 1, D-14476 Potsdam, Germany
 }

\date{}

\begin{document}

\maketitle

\begin{abstract}
\noindent

We undertake first steps in making a class of discrete models of quantum gravity, spin foams, accessible to a large scale analysis by numerical and computational methods. In particular, we 
apply Migdal-Kadanoff and Tensor Network Renormalization  schemes to
spin net and spin foam models based on finite Abelian groups and introduce `cutoff models' to probe
the fate of gauge symmetries under various such approximated renormalization group flows. For the
Tensor Network Renormalization analysis, a new Gau\ss~constraint preserving algorithm is introduced
to improve numerical stability and aid physical interpretation. We also describe the fixed point
structure and establish an equivalence of certain models.

\end{abstract}

\tableofcontents


\section{Introduction} \label{intro}


Spin foam models aim at providing a description of the microscopic structure of spacetime and thus a
theory of quantum gravity
\cite{Reisenberger:1994aw,Reisenberger:1996pu,Barrett:1997gw,Baez:1997zt,Perez:2003vx,Livine:2007vk,
Engle:2007uq,eprl,Freidel:2007py,Dupuis:2011fz,Bahr:2011yc}.
These models can be understood as a non--perturbative definition of the path integral for quantum gravity.
To make these path integrals well defined one has to introduce a regularization based on a choice of
discretization, i.e.\ a lattice or, more generally, a triangulation or two--complex. Indeed, spin
foams can be understood as generalized lattice gauge theories.

This discrete (a priory auxiliary) structure should not be confused with another feature of spin foams, which is often termed  `Planck scale discreteness' \cite{Rovelli:1994ge,Ashtekar:1996eg,Ashtekar:1997fb,Dittrich:2007th,Rovelli:2007ep}, namely that the spectra of (kinematical) geometrical quantum observables, like areas and volumes, are discrete. There are thus two different kinds of UV cutoffs, whose interplay has not been fully understood yet. This has to be kept in mind when discussing a possible breaking of (global) Lorentz or (local) diffeomorphism symmetry. A (naive) lattice regularization will generically break these symmetries, see for instance \cite{Loll:1997iw,Gambini:2002wn,Gambini:2005sv,Dittrich:2008pw,Bahr:2009ku,Bahr:2009mc,Gambini:2011nx,Polchinski:2011za} for a discussion of these issues in gravity. 

We may, however, consider a continuum limit with respect to this auxiliary discretization scale, for example by a coarse graining or blocking procedure, see \cite{Bahr:2009qc,Bahr:2010cq,Bahr:2011uj} for recent examples involving gravity or related to it. 
A crucial question then is whether Lorentz or diffeomorphism symmetry will be restored in this limit, despite the possibility of still having the second kind of UV cutoff, provided by the discreteness of the spectra, in the theory. That this cutoff does not necessarily lead to a violation of Lorentz symmetry has, for instance, been argued in \cite{Rovelli:2002vp} on kinematical grounds.
 A full dynamical scenario for 4D gravity where a restoration has been shown to occur is, however,
missing, nonetheless see
\cite{Gambini:2008as,Gambini:2008ea} for progress in this direction.\footnote{For 3D gravity, which is a topological theory, that is without propagating degrees of freedom, discretization does not necessarily lead to a breaking of diffeomorphism symmetry \cite{Freidel:2002dw,Noui:2004iy,Dittrich:2008pw}. This holds also for 3D gravity with cosmological constant \cite{Turaev:1992hq}. One can, however, consider discretization or quantization methods which a priori break diffeomorphism symmetry and look for methods to restore these symmetries, see \cite{Bahr:2009qc,Perez:2010pm,Noui:2011im}.}

These questions motivate us to consider a continuum limit which involves many building blocks or large lattices (with many vertices), as this is the limit where one can hope to obtain a diffeomorphism invariant theory. An alternative is the semi--classical limit \cite{Barrett:2009gg,Barrett:2010ex,Conrady:2008mk}, in which rather  the Planck constant, leading to `Planck scale discreteness', is taken to zero. To distinguish these two kind of limits we will sometimes refer to the first one as statistical limit.

Experience with other quantum gravity models, such as (causal) dynamical triangulations, has shown
\cite{Loll:1998aj,Ambjorn:1995jt,Ambjorn:1997di,Ambjorn:2005jj,Ambjorn:2011ph,Ambjorn:2011cg,
konopka} that even before the question of restoration of symmetries can be addressed, it is not  at
all obvious whether such a statistical limit leads to any viable model of spacetime, i.e.\ whether
such a limit will result in smooth four--dimensional spacetime manifolds. Indeed, in this kind of
limit statistical considerations become important and it can easily happen that state sums become
entropically dominated by configurations not resembling any four--dimensional manifold at all. As we
will see, a related issue arises for spin foam models (or other models based on first order/tetrad
formulations) where geometrically degenerate configurations might be dominant. Such configurations
also turn up in a semi--classical or classical phase space analysis, even if this involves only a
single simplex \cite{Dittrich:2008ar,Dittrich:2010ey,Barrett:2009as}.

Hence it is crucial to investigate which kind of large scale physics or, in other words, phases, are
encoded in the candidate quantum gravity models. Phases are often characterized by symmetries, that
is such a study might also answer which kind of symmetries might be restored in a large scale~/
statistical limit. Indeed, making progress in this direction is one of the most pressing issues for
the spin foam approach. However, it is also a long standing open issue
\cite{Markopoulou:2000yy,Markopoulou:2002ja,Oeckl:2002ia} charged with a number of conceptual and
technical challenges.

One important challenge is the complex structure of the models which lead to very complicated
amplitudes as compared, for instance, to QCD. Here, our strategy \cite{Bahr:2011yc} is to develop a
wide range of simplified models which capture essential features of spin foams while being much
easier to handle. These simplifications are obtained on the one hand by replacing Lie groups, on
which the gravitational spin foams are based, with finite groups. On the other hand, we can also
consider `dimensionally reduced' models  (spin net models). In fact, 2D spin net models of the
simplest class share many statistical properties with their counterpart 4D spin foam models.

Similar simplified models have been successfully studied, e.g.\
\cite{Creutz:1979kf,Creutz:1979zg,Creutz:1982vz}, to get insights into the large scale behavior of
lattice gauge theories. Here we hope for a similar improved understanding of the possible phases
that can occur in quantum gravity models. In particular, we will see in the course of the paper, how
conjectures or even conclusions for Lie groups can be made based on findings for finite groups.
These simplified models can also be interesting in their own right
\cite{Liberati:2009uq,Gu:2009jh,Hamma:2010sc}, in particular if an example is found in which some
analogue to diffeomorphism symmetry is restored. Indeed topological phases and  string net
condensates \cite{Levin:2004mi} which are studied in condensed matter, also regarding the question
of symmetry restoration, are tightly related to 3D spin foams with finite groups \cite{Bahr:2011yc}.

The development of coarse graining and renormalization techniques seems to be the most promising
avenue to study the large scale behavior and simplified models allow us to adapt and further
refine methods from lattice gauge theory and condensed matter systems. In this work we will
therefore apply the Migdal-Kadanoff scheme \cite{Migdal:1975zg,Kadanoff:1976jb} and the tensor
network renormalization (TNR) method \cite{levin,Gu:2009dr}. These schemes involve a regular lattice
and, due to this regular structure, they are amenable to efficient numerical simulations. 

With this approach we are able to explicitely answer the question of $BF$ symmetry restoration for a range of models and to gain insights into how these results are related to the case of infinite groups. These results should be understood as a first step towards harnessing the power of numerical methods from statistical physics to deepen the understanding of the large-scale physics of spin foam models.



In this work, we will first introduce spin foam and spin net models and write them in ways suitable for coarse graining (section \ref{spin net}). 
We will also define a particularly important class of models, termed `Abelian cutoff models', discuss the role of $BF$ / translation symmetry and detail the relationsship between spin foams and nets. 
Subsequently, in section \ref{coarsemethods}, we discuss the conceptual challenges of coarse graining in this context and argue for the approach pursued here, in particular for the use of a regular lattice.

We then apply the Migdal-Kadanoff and tensor network approximation schemes to coarse grain our models (sections \ref{mk} and \ref{tnr}, resp.). In each case, we first introduce the method and highlight some of its analytical properties before presenting numerical results that focus on the question whether renormalization of Abelian cutoff models will restore $BF$ symmetry. In particular, we feature a Gau\ss~constraint preserving TNR algorithm tailored to the geometric interpretation of our models and establish an equivalence amongst certain cutoff models under the TNR renormalization scheme. We conclude by comparing both approximation schemes and pointing out possible future directions of research.


\section{Spin foams and spin net models} \label{spin net}

Spin foams are a particular class of lattice gauge models (see e.g. \cite{Alejandro} for a recent
review and \cite{Bahr:2011yc} for a review emphasizing the relation to lattice gauge  and
statistical
physics models). Such models are specified by variables, taking values in some group $G$, 
associated to the edges of a lattice (or more generally an oriented 2--complex) and weights
associated to the plaquettes. They can thus also be termed plaquette models.

A related class of models, which will be introduced below, are so called edge or spin net models
\cite{Bahr:2011yc}. Here group variables are associated to the vertices of a lattice (or more
generically an oriented graph or 1--complex) and weights to the edges. This class includes the
well--known Ising models, based on the group $\Zl_2$. Indeed it will turn out that the structures
involved in a spin net model are very similar to those involved in spin foam models -- just that
where, for instance, weights are associated to 2D plaquettes for spin foams, weights are associated
to
1D edges in spin nets, similarly for the group variables and so on. In this sense spin nets are a
simpler or dimensionally reduced form of spin foams. There is however one essential difference,
which is
that spin foams enjoy a local gauge symmetry whereas spin nets only feature a global symmetry, in
both cases given by the group $G$ the models are based on. In section 
\ref{subsec:relation} we will also comment on another relationship between spin foams and spin
nets: spin nets can be seen as measuring non gauge-invariant observables in spin foam models.

Spin foams and spin nets are defined by partition functions, and we will first consider a
representation of these partition functions as sums over group variables. Via a group Fourier
transform we can rewrite these partition functions as sums over variables labeling the irreducible
representations of the group $G$. This is where the name 'spin foam' stems from, as 'spin' refers to
the representation labels for the group $SU(2)$. This alternative representation is well known as a
duality transformation for both edge and plaquette models, and is usually employed for the high
temperature or strong coupling expansion \cite{Savit:1980}.  The models in this representation are
not only specified by the dual weights but also by an intertwining projector acting on a certain
representation space. Non--trivial spin foam or spin net models can be constructed by choosing this
projector to be different from its standard form (which is the Haar intertwiner introduced below) in
plaquette and edge models respectively. In the case of Abelian groups we will explicitly construct
non-trivial models, the so-called \emph{Abelian cutoff models}, which only consider representation
labels of the group up to a certain cutoff. For these models, we will see that the choice of a
non-trivial projector is equivalent to retaining the Haar intertwiner while restricting the
dual weights in a particular way. These are precisely the models that we will numerically analyze in
sections \ref{mk} and \ref{tnr}.

For the non-trivial models, one can then apply the inverse group Fourier transform and again obtain a partition function in terms of group variables. As will be explained below, this will however require the introduction of several group variables per vertex (for edge models) or edge (for plaquette models) \cite{Pfeiffer:2001ig,Pfeiffer:2001ny,bahrta}. This representation is termed holonomy representation, both for spin foams and spin nets.

In the next two subsections we will give a short introduction into the main concepts and different
representations of spin foams and spin nets. Furthermore we will detail the different possibilities
of rewriting these models into tensor network form, as this will be the basis of one of our coarse
graining methods, to be discussed later on.

\subsection{Spin net models} \label{spinnets}

To construct  spin net models we start with state sum models formulated over a finite group $G$ and on a graph (one-dimensional
complex) with oriented edges. More precisely we consider partition functions of the type 
\be\label{Z-hol1}
 Z=\frac1{|G|^{\sharp v}}\sum_{\{g_v\}}\prod_e w_e(g_{s(e)}g^{-1}_{t(e)}),
\ee
where $\sharp v$ is the number of vertices in the graph, $s(e)$ denotes the source
vertex (starting point) of the edge $e$, $t(e)$ denotes its target vertex (final point), and the curly
brackets under the sum symbol denote that there is a sum per vertex:
$\sum_{\{g_v\}}=\prod_v\sum_{g_v}$.
Here group elements are associated to vertices and weights, which determine the couplings, to edges.
Therefore these models are also known as edge models and include the standard Ising model for which
$G$ is equal to $\Zl_2$.
The weights $w_e(g_{s(e)}g^{-1}_{t(e)})$ can be arbitrary functions\footnote{To have a statistical interpretation of $w_e$ as probability weights these should be positive. This will however not necessary be the case for spin foam models.}  over the group $G$. However, if $w_e$ are class functions, i.e.\ invariant under conjugation ($w_e(g)=w_e(hgh^{-1})$ $\forall g,h\in G$), the model will feature a global symmetry given by the group $G$: the partition function remains invariant when applying the same conjugation to group variables at each vertex.

The form (\ref{Z-hol1}) defines the simplest form of spin net models in the representation based on group variables. This representation will be called holonomy representation in analogy to spin foam models (where group variables are associated to edges and represented holonomies of a connection). Holonomy representations of more general models will require several group variables associated to each vertex, as we will see later.

Via the group Fourier transform, we can change from the above representation in terms of group elements, to the \emph{spin net} representation, which is in terms of the irreducible representations of the group\footnote{We refer the reader e.g. to \cite{Fulton} for the main concepts of group representation theory that we employ.}.
Every function on the group can be decomposed in terms of matrix elements of  the irreducible representations $\rho$, 
\be\label{p-w}
  w(g)=\sum_\rho\sum_{a,b=1}^{\text{dim}\rho}\,(\tilde w_\rho)_{ab}\,\rho(g)_{ab},\qquad
 (\tilde w_\rho)_{ab}=\frac{\text{dim}\rho}{|G|}\sum_g\,w(g)\,\rho^*(g)_{ab},
 \ee
being $\rho^*$ the dual of $\rho$.
For class functions this decomposition reduces to the character decomposition
\be
 w(g)=\sum_\rho \,\tilde w_\rho\,\chi_\rho(g),\qquad
\tilde w_\rho=\frac1{|G|}\sum_g\,w(g)\,\chi_{\rho^*}(g),
\ee
where $\chi_\rho(g)=\sum_{a=1}^{\text{dim}\rho} \rho(g)_{aa}$ denotes the character. We note
that our convention for the delta function over the group, $\delta_G$, is 
\be
\frac1{G}\sum_g\delta_G(g)f(g)=f(\text{id}),\qquad \delta_G(g)=\sum_\rho
\text{dim}\rho\,\chi_\rho(g).
\ee

\begin{figure}
\begin{center}
\includegraphics[width=0.9\textwidth]{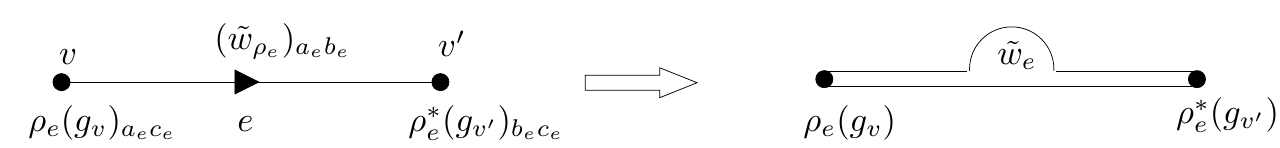}
\caption{On the left: the three objects associated to every edge. On the right: their schematic
representation. Every straight line joining two objects means a contraction of
indices. 
\label{fig:edge}}
\end{center}
\end{figure}

Using the property
\be\label{properties}
 \rho(g^{-1})_{ab}=\rho^*(g)_{ba}.
\ee
 we obtain
\be\label{Z-sn-g}
 Z=\frac1{|G|^{\sharp v}}\sum_{\{g_v\}}\sum_{\{\rho_e\}}\prod_e
\,(\tilde w_{\rho_e})_{a_eb_e}\, \rho_e(g_{s(e)})_{a_ec_e}
\,\rho^*_e(g_{t(e)})_{b_ec_e},
\ee
where we sum over repeated indices.\footnote{For class functions $w_e$ we have
$(\tilde w_{\rho_e})_{a_eb_e}=(\tilde w)_{\rho_e}\,\delta_{a_eb_e}$, which will contract the
representation matrices $(\rho_e)_{a_ec_e}$ to the characters $\chi_{\rho_e}$.} 
 Note that, associated to every edge, there is a
coefficient $(\tilde w_{\rho_e})_{a_eb_e}$, and two group representations,
$\rho_e(g_{s(e)})_{a_ec_e}$ living on the source vertex and $\rho^*_e(g_{t(e)})_{b_ec_e}$ living in
the target vertex. The indices of these three objects are contracted, as described schematically in
figure \ref{fig:edge}.

\begin{figure}
\begin{center}
\includegraphics[width=0.8\textwidth]{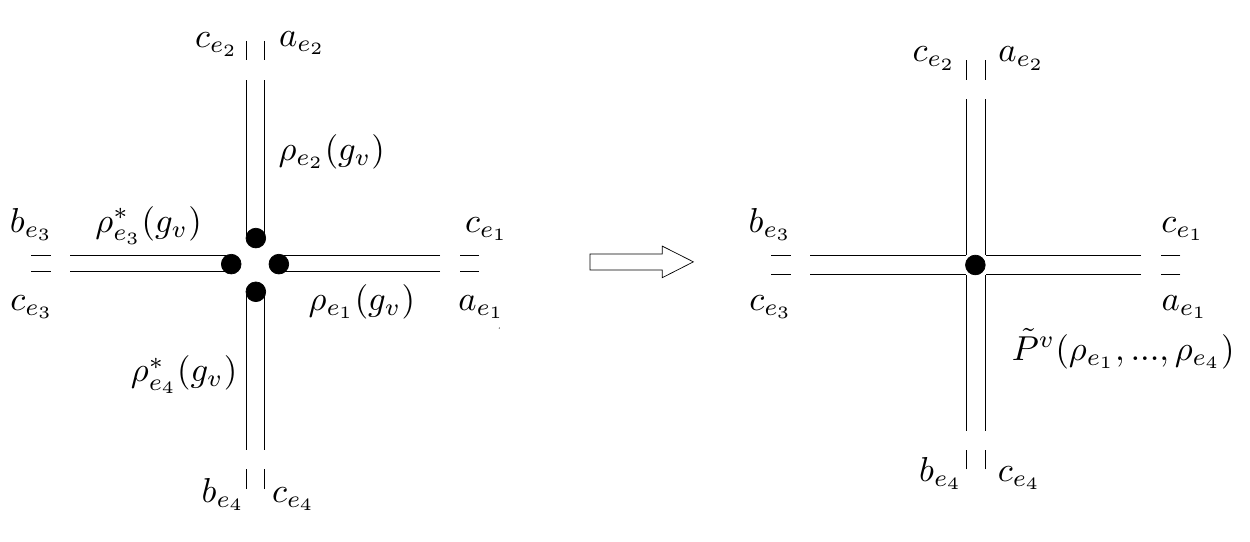}
\caption{Four-valent vertex with two outgoing edges, $e_1$ and $e_2$, and with two incoming edges,
$e_3$ and $e_4$. On the left: representations meeting in the vertex. On the right: schematic
representation of the resulting vertex weight.
\label{fig:vertex}}
\end{center}
\end{figure}

Now we can carry out the sums over group variables. The result is
\be\label{Z-spin}
 Z=\sum_{\{\rho_e\}}\big(\prod_e(\tilde w_{\rho_e})_{a_eb_e}
\big)
\prod_v \tilde{P}^v_{a_e...,b_{e'}...;c_e...,c_{e'}...}(\{\rho_e\}_{e\supset v})
\ee
where we have defined the vertex weight
\begin{align}\label{V-spin}
\tilde{P}^v_{a_e...,b_{e'}...;c_e...,c_{e'}...}(\{\rho_e\}_{e\supset
v}):=\frac1{|G|}\sum_{g_v}\prod_{e:{v=s(e)}}
\rho_e(g_v)_{a_ec_e}\;\prod_{e':{v=t(e')}}\rho^*_{e'}(g_v)_{b_{e'}c_{e'}}.
\end{align}
The first and third groups of indices involve all the edges for which $v$ is the source vertex. The
second and fourth groups of indices involve all the edges for which $v$ is the target vertex.
Using the schematic representation employed in figure \ref{fig:edge}, we can
represent the vertex weight as in figure~\ref{fig:vertex}.

Note that $\tilde{P}^v$ can be seen as an intertwiner map, called the Haar intertwiner, acting
on a certain representation space for the group $G$. This representation space, ${\cal H}_v$,
associated to the vertex $v$, is given by the tensor product of  the representations $\rho_e$
associated to the outgoing edges and the representations $\rho^*_{e'}$ associated to the incoming
edges 
\ba\label{tprod}
\tilde{P}^v:   \left( \bigotimes_{e:{v=s(e)}}  \!V_{\rho_e} \right)\otimes    \left(
\bigotimes_{e':{v=t(e')}}  \!V_{\rho^*_{e'}} \right)   \,\rightarrow\,  
 \left( \bigotimes_{e:{v=s(e)}}  \!V_{\rho_e} \right)\otimes    \left( \bigotimes_{e':{v=t(e')}} 
\!V_{\rho^*_{e'}} \right).
\ea
The intertwining property of this map is guaranteed by the averaging over the group in
(\ref{V-spin}). Indeed, the Haar intertwiner defines an orthogonal  projector onto the subspace
${\cal H}_v^{inv}$ of ${\cal H}_v$, invariant under the group action defined on this representation
space.

The same intertwining map will appear in spin foam models. There, the choice of the Haar intertwiner
as a projector and face weights to be trivial defines topological models, known as $BF$--theories. A
gauge symmetry, known as translation symmetry, arises in this case, forcing the model not to have
local physical degrees of freedom. The analog situation happens in spin net models. The choice of
edge weights  $\tilde w_e \equiv \text{id}$ and of the Haar intertwiner as a projector corresponds
to the model at zero temperature, with no local degrees of freedom. For the weights $w_e$
this amounts to choosing $w_e\sim \delta_G$.  In this case the projectors $\tilde{P}^v$
are just contracted along the edges of the graph.

Non-trivial models are more interesting. They are constructed  by restricting the projector further,
i.e.\ by selecting a subspace of the invariant subspace of ${\cal H}_v$ and by replacing the Haar
intertwiner (\ref{V-spin}) by a projector onto this subspace.
We will proceed in that way here, and in general assume that $\tilde{P}^v$
is a projector onto some subspace of ${\cal H}_v^{inv}$.
This allows us to obtain interesting models even if we choose the edge weights $\tilde
w_e$ to be trivial. Indeed, in these models, the original gauge symmetry is broken and local physical degrees of freedom arise. We will see this behavior for the Abelian cutoff models described below.

In general, in order to re--express the partition function $Z$ of such non-trivial models in the holonomy representation, namely as a sum over group variables, we will need to associate more than one group elements to each of the vertices in the lattice.
Let us denote the number of edges attached to the vertex $v$ by $n_v$. Now, we assign one group
element $g(v,e)$ to each of the edges attached to $v$ and define 
\begin{align}\label{V-g}
P^v(\{g_{(v,e)}\}_{e\supset v})&=
\sum_{\{\rho_e\}}\tilde{P}^v_{a_e...,b_{e'}...;c_e...,c_{e'}...}(\{\rho_e\}_{e\supset v})\nonumber\\
&\times
\prod_{e|v=s(e)}\text{dim}(\rho_e)\;\rho^*_e(g_{(v,e)})_{a_e c_e}\;
\prod_{e'|v=t(e')}\text{dim}(\rho_e')\;\rho_{e'}(g_{(v,e')})_{b_{e'}c_{e'}}\q .
\end{align}
In terms of this vertex amplitude, the partition function reads
\begin{align}\label{Z-hol2}
Z&=\big(\prod_v\frac1{|G|^{n_v}}\big)\sum_{\{g_{(v,e)}\}}\big(\prod_ew_e(g_{(s(e),e)}
g^{-1} _{(t(e),e)})\big)\big(\prod_vP^v(\{g_{(v,e)}\}_{e\supset v})\big).
\end{align}
In case that $\tilde P^v$ is given by (\ref{V-spin}), i.e.\ by the Haar intertwiner, we obtain
the form of the partition function given in (\ref{Z-hol1}), as $P^v$ then enforces equality between
the group
elements associated to one and the same vertex. 

A different simplification occurs when $w_e(h)\sim\delta_{G}(h)$, a choice that we already mentioned
above. Then the two group elements $g_{(v,e)}$ and $g_{(v',e)}$ associated to any edge $e$ have to
be equal. Hence, the partition function reduces to a sum over group elements
$g_e=g_{(v,e)}=g_{(v',e)}$ associated to edges. This type of models is known as vertex models -- the
energy of a configuration is now determined by the vertex weights $P^v(\{g_e\}_{e\supset v})$.

In this work we will apply coarse graining to models with Abelian groups $\Zl_q$. In this case, all
irreducible unitary representations, which we will label by $k \in \Zl_q$, are one--dimensional and
defined by their characters $\chi_k(g)=\exp(\tfrac{2\pi i}{q} k\cdot g)$ for $g \in \Zl_q$. The
transformation between functions on the group $\Zl_q$ and on the dual, equal to the space of
characters, which is also given by $\Zl_q$, is given by the discrete Fourier transform
\be\label{za3}
w(g) = \sum_{k=0}^{q-1}   \tilde w_k \, \chi_k(g)  \qquad
\tilde w_k = q^{-1} \sum_{g=0}^{q-1}  w(g)  \overline{\chi}_k(g)\, .
\ee
Characters for Abelian groups are multiplicative, i.e.\ $\chi_k(g_1\cdot g_2)=\chi_k(g_1) \cdot
\chi_k( g_2)$, and also $\chi_k(g^{-1})=\chi_k^{-1}(g)=\overline{\chi}_k(g)$. Moreover, the
delta over the group is now the $q$-periodic delta. It verifies
$q^{-1}\sum_g\delta^{(q)}(g)f(g)=f(0)$.

The spin net representation of the partition function simplifies in the case of Abelian groups to
 \be \label{Z-sn-abelian}
Z= \sum_{\{k_e\}}   \prod_e \tilde w_{k_e}  \prod_v \tilde{P}^v(\{k_e\}_{e\supset v}),\qquad
\tilde{P}^v(\{k_e\}_{e\supset v}):=\prod_{e\supset v} 
\delta^{(q)} \!\big( \sum_{e\supset v} \varep^e_v k_e\big),
\ee
where $\varep^e_v$ is equal to $+1$ ($-1$) if $v$ is the source (target) of $e$. 
The projector $\tilde{P}^v$ implements the Gau\ss~constraints at the vertex $v$. It indeed projects to the irreducible subrepresentation in the tensor product of all representations associated to the outgoing edges and the tensor product of dual representations of incoming edges. The tensor product is one--dimensional and equal to the trivial representation if the oriented sum of the representation labels $k_e$ is equal to zero. 

Note that the spin net representation is the starting point for the high temperature expansion
\cite{Oitmaa:2006}. The infinite or high temperature fixed point is represented by $\tilde w_e(k)
=\delta^{(q)}(k)$. On the other hand the zero or low temperature fixed point is given by $\tilde
w_e(k)\equiv 1$. This corresponds to weights $w_e(h)=\delta_G(h)$ in the original group
representation. The model is `frozen', i.e.\ the group elements at
the different vertices have to agree (assuming that the graph has only one connected component).

As commented before, for this choice of weights there is a gauge symmetry. It is  associated to the
faces, i.e.\ the two--dimensional cells (here we are assuming that the graph is actually given by an
orientable 2--complex, i.e.\ the 2--dimensional cells are well defined). Associating to every face
$f$ an element $k_f \in \Zl_q$ we define a gauge transformation acting as 
\ba\label{transisn}
k_e \mapsto k'_e=k_e + \sum_{f \supset e} \varepsilon^f_e k_f
\ea
where $\varepsilon_e^f$ is $+1$ ($-1$) if the orientations of $e$ and $f$ agree (disagree). Under
such a transformation the contribution of a configuration $\{k_e\}_e$ to the partition function $Z$
does not change.
Choosing either the
edge weights $\tilde w_e$ or the vertex projector $\tilde P^v$ to be non--trivial will in
general break this translation symmetry either completely or down to a smaller symmetry.
Choosing a non-trivial projector $\tilde P^v$ will in general result in vertex models, as one can
basically reduce the set of vertices allowed by the Gau\ss~constraints even further.

We are now in position to introduce the Abelian cutoff models.
As said before, for the low temperature fixed point (analog to BF in spin foam models) the weights are
\ba
  w_e(g) = \delta^{(q)} (g) = \sum_{\{k\}} \chi_k (g)\quad \leftrightarrow\quad \tilde w_e(k)=1\, \forall k.
\ea
The Abelian cutoff models are derived from this by `cutting off' the sum at some value $K$ such that some of the dual weights $\tilde w_e$ vanish. Explicitly,
\ba\label{Aco}
\tilde w_e(k) =
\begin{cases}
&1,\q\text{for}\; |k|\leq K   \\
&0,\q\text{for}\;|k|>K  \q 
\end{cases}
\ea
where we will consider only even $q$, hence $K \leq \frac{q}{2}$. Here the range for $k$ is given
by $\frac{-q}{2}<k\leq\frac{q}{2}$.
Also, note that the symmetry condition $\tilde{w}(k) = \tilde{w}(-k) \, \forall k$ is fulfilled.
This requirement is desired in the quantum gravity setting because it ensures that the model does not depend on edge orientations and certain types of face and edge subdivisions \cite{Bojowald:2009im,Bahr:2010my}.

Abelian cutoff models could be equivalently described by a restriction
of the projector $\tilde P^v$ (and keeping the weights $\tilde w_e(k)\equiv 1$).
They provide a simple example of breaking the translation gauge symmetry from the frozen model in
order to introduce physical degrees of freedom. Also this choice of model corresponds to a
regularization one often choses for $BF$ lattice theories with Lie groups \cite{PR}. In this case
the orbits of the translation gauge symmetry are non--compact and the evaluation of the partition
function gives generically infinity. Different methods of regularization have been developed, one
would be equivalent to introducing a cutoff $K$ (for a theory with group $G=U(1)$).  One can then
ask whether these regularized models would flow back to the full $BF$ model, or more generally the
low temperature fixed point, under coarse graining. We will consider this question in sections
\ref{mk} and \ref{tnr}.


So far we have presented the partition functions for spin net models as sums over group variables
(the holonomy representation) or representation labels (the spin net representation). An alternative
is to write the partition function as a contraction over tensors attached to the vertices of the
underlying lattice (or some graph associated to the lattice), that is, in the tensor network
representation. The tensor network representation is commonly employed in statistical and quantum
systems \cite{vidalprl,Singh:2009cd,singh2,
levin,Cirac}, since it is especially suitable for developing techniques of renormalization. We will
make use of this representation of the partition function in section \ref{tnr}, where we will apply
the renormalization approach to spin net models with Abelian groups.

In this representation, the contraction of the indices is prescribed by the edges.
The partition function can hence be expressed as a tensor trace
\ba\label{Z-tn}
 Z= \sum_{\{k_e\}}\prod_v \tilde{T}^v_{\{k_e\}_{e\supset
v}}\equiv\text{tTr}\otimes_v\tilde{T}^v.
\ea
This way of representing the partition function is related to the so--called  graphical calculus
\cite{penrose1,penrose2,biamonte,Barrett:2009mw}, which is often employed in the spin foam
literature.

\begin{figure}
\begin{center}
\begin{overpic}
[width=0.5\textwidth]{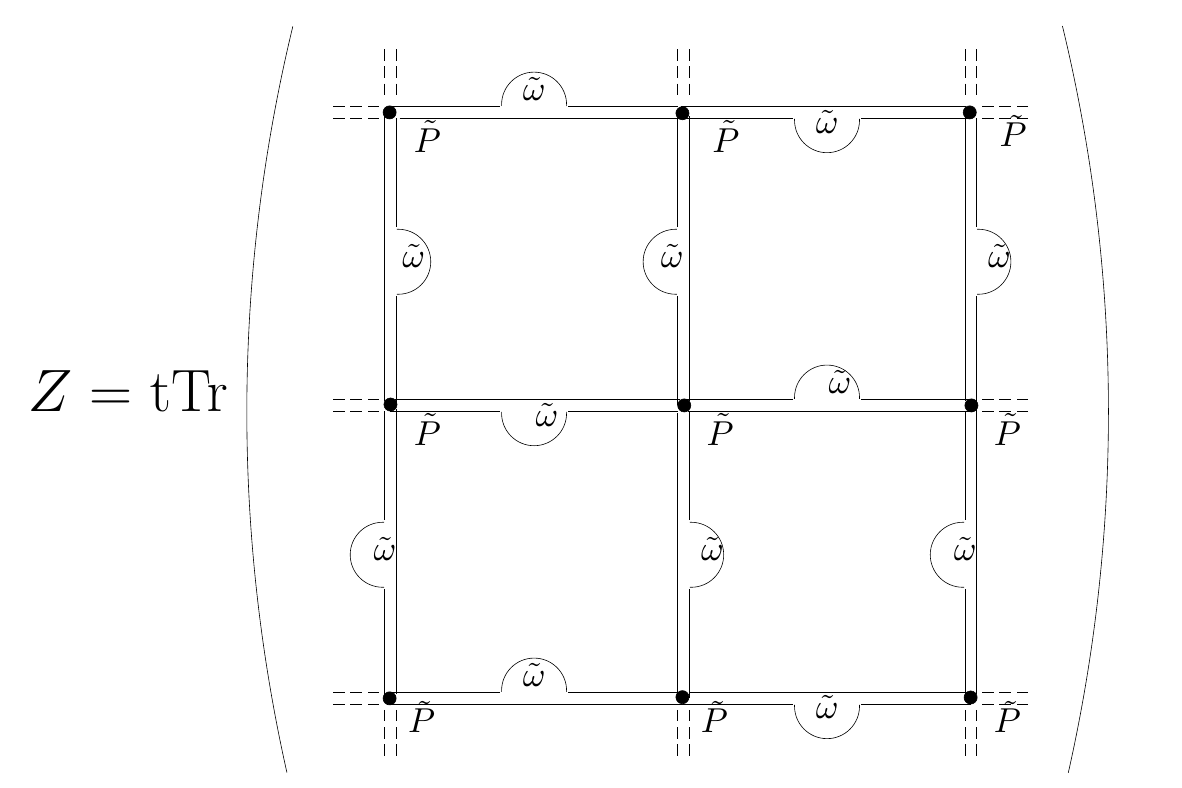}
\end{overpic}
\caption{
Employing the schematic representation of figures \ref{fig:edge} and \ref{fig:vertex} it is
straightforward to realize that the partition function of spin net models can be written in
the form of a tensor-trace over a network of tensors, being the tensor-trace the sum over
representations.
\label{fig:tnw1}}
\end{center}
\end{figure}

\begin{figure}
\begin{center}
\begin{overpic}
[width=0.7\textwidth]{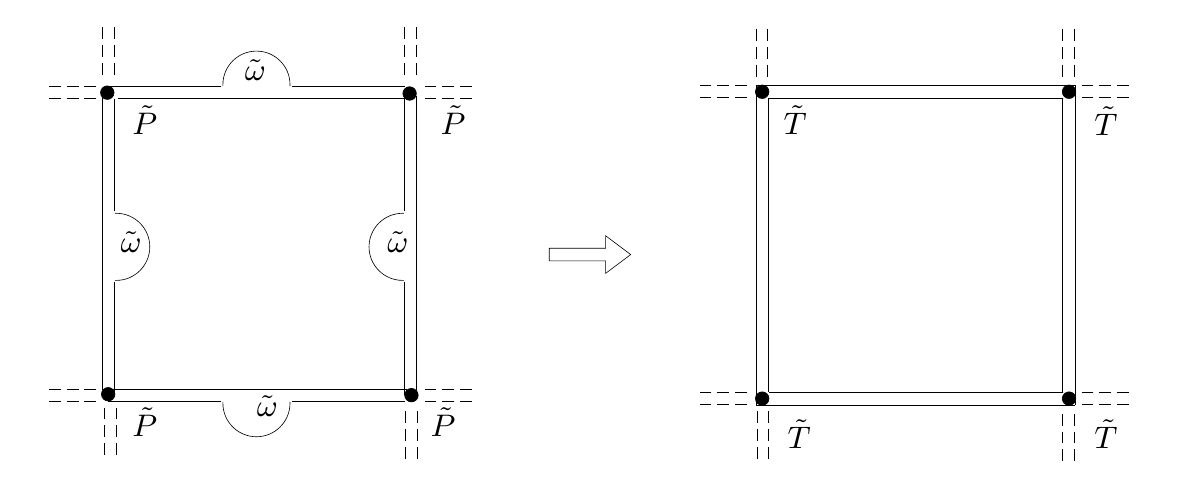}
\end{overpic}
\caption{Schematic definition of the vertex tensor $\tilde{T}^v_{\{k_e\}_{e\supset v}}$ in a
square lattice.
\label{fig:tnw2}}
\end{center}
\end{figure}

In the particular example of Abelian models in the spin net representation,
we first absorb the edge weights $\tilde w_{k_e}$ into the vertex
weights $\tilde{P}^v(\{k_e\}_{e\supset v})$ by distributing $\tilde w_{k_e}^{1/2}$ factors to
each of the adjacent vertices, namely in every vertex we define the tensor 
\be
\tilde{T}^v_{\{k_e\}_{e\supset v}}=\big(\prod_{e\supset v}
(\tilde
w_{k_e})k^{1/2}\big)\tilde{P}^v(\{k_e\}_{e\supset
v})
\ee
The partition function is then given by the tensor trace (\ref{Z-tn}) (see figure
\ref{fig:tnw1} and figure \ref{fig:tnw2}).

More generally, we can find tensor network representations also for the non--Abelian models in the
different representations. The holonomy representations is a convenient starting point for a low
temperature expansion, the spin net representation to the high temperature region. For the spin net
representation we can proceed as for the Abelian models. For the holonomy representation assume that
we have edge weights such that $w_e(h)=\delta_G(h)$. We then just need to understand the group
elements $g_e$ as indices attached to the tensors $T^{s(e)}=P^{s(e)}$ and $T^{t(e)}=P^{t(e)}$ to see
that (\ref{Z-hol2}) can be rewritten as a tensor trace. For the more general case of non--trivial
edge weights we can introduce another set of rank two tensors $w$ to the midpoints of the edges.

\subsection{Spin foam models} \label{spinfoams}

Spin foams are in many respects similar to spin nets. The main difference is that spin foams are gauge theories, formulated with a gauge group $G$, which here will be a finite group. Furthermore spin foams require an oriented two--complex for their definition. This implies, that we have a well defined notion of oriented edges as well as oriented faces (which are the 2D cells of the complex, or  the plaquettes for a regular lattice). 

In order to define a spin foam model, we assign a group element $g_e$ to every edge $e$ of the two-complex and a weight $w_f:G\rightarrow\mathbb{C}$ to every face $f$. The state sum model is then defined by the partition function
\be\label{Z-SF-hol1}
 Z=\frac1{|G|^{\sharp e}}\sum_{\{g_e\}}\prod_{f} w_f(h_f),
\ee
where $\sharp e$ denotes the total number of edges in the two-complex.

\begin{figure}
\begin{center}
 \mbox{
      \subfigure[definition of curvature]{\includegraphics[width=0.25\textwidth]{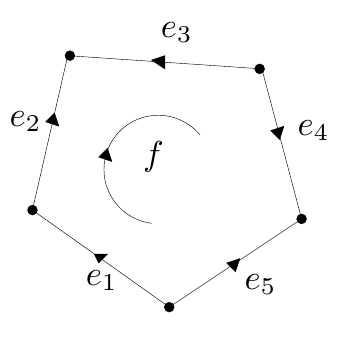}} \qquad \qquad \qquad
      \subfigure[edge and faces attached to
it]{\includegraphics[width=0.3\textwidth]{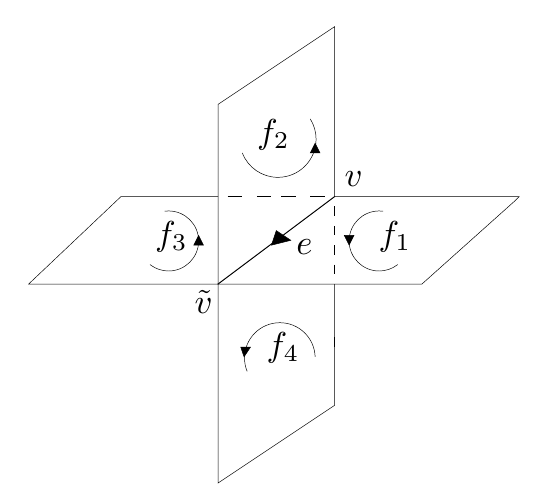}}
      }
\caption{($a$) A face $f$ bound by the edges $e_1,\cdots,e_5$. The curvature is
$h_f=g_{e_1}g_{e_2}g_{e_3}^{-1}g_{e_4}g_{e_5}^{-1}$.\qquad
($b$) An edge $e$ with four faces attached, $f_1$ and
$f_4$ with
 positive relative orientation with respect to $e$, and $f_2$ and $f_3$ with negative relative
 orientation. The corresponding edge-weight is denoted by
$\tilde{P}^e_{a^{f^1}_{v},a^{f^4}_{v},a^{f^2}_{v},a^{f^3}_{v};
a^{f^1}_{\tilde v},a^{f^4}_{\tilde v},a^{f^2}_{\tilde v}, a^{f^3}_{\tilde v}}$.
}\label{fig:curvature}
\end{center}
\end{figure}

The function $w_f$ is a class function.
Furthermore, in \eqref{Z-SF-hol1} $w_f$ depends on the group elements only through the
holonomy $h_f$ around the closed loop of edges forming the face $f$, that we will call
\emph{curvature}. Let us make its definition explicit.
For that, we recall that the relative orientation between a face $f$ and any edge $e$ in the
boundary of $f$ is denoted by $\varepsilon^f_e$, and it is equal to $+1$ ($-1$) when $e$ and
$f$
have the same (opposite) orientation. Given a face $f$ bounded by the ordered sequence of edges
$e_1,e_2,\cdots,e_n$, the associated curvature is given by
\be
h_f=g_{e_1}^{\varepsilon^f_{e_1}}g_{e_2}^{\varepsilon^f_{e_2}}\cdots
g_{e_n}^{\varepsilon^f_{e_n}},
\ee 
as depicted in figure \ref{fig:curvature}$a$.
These properties of the weight $w_f$ guarantee that the partition function is invariant under gauge
transformations $g_e\rightarrow g_{s(e)}g_eg_{t(e)}^{-1}$. As before, $s(e)$ denotes the
source vertex of the edge $e$ while $t(e)$ denotes its target vertex. 

The partition function (\ref{Z-SF-hol1}) describes standard lattice gauge theories. The weights can
for instance be chosen to emulate the Wilson action
\be
w_f(h)\; = \; \exp(-S_W(h)) \q , \q\q S_W\;=\;\frac{1}{2\alpha} (\chi_\rho(h) + \chi_\rho(h^{-1}))
\q ,
\ee
where $\alpha$ is a coupling constant. 
For the choice $w_f(h_f)=\delta_G(h_f)$, with $\delta_G$ the delta function over the group
defined as before, the partition function only sums over (locally) flat holonomies. This is a
discretization of $BF$ theory, which is a  topological field theory (without propagating
degrees of freedom). It coincides with the zero temperature fixed point or zero coupling fixed point
of lattice gauge theories of Yang Mills type.

In order to obtain a representation of the partition function (\ref{Z-SF-hol1}) as a sum over representation labels (spin foam representation), we again apply the group Fourier transform. 
Here we only need to decompose class functions into characters
\be
 w(g)=\sum_\rho \,\tilde w_\rho\,\chi_\rho(g),\qquad
\tilde w_\rho=\frac1{|G|}\sum_g\,w(g)\,\chi_{\rho^*}(g) \q .
\ee
To decompose $w_f(h_f)=w_f(g_{e_1}^{\varepsilon^f_{e_1}}g_{e_2}^{\varepsilon^f_{e_2}}\cdots
g_{e_n}^{\varepsilon^f_{e_n}})$
we use the properties
\be\label{properties2}
\chi_\rho(gh)=\sum_{ab}\rho(g)_{ab}\rho(h)_{ba},\qquad \rho(g^{-1})_{ab}=\rho^*(g)_{ba}.
\ee
We  introduce this decomposition in equation  \eqref{Z-SF-hol1} and individually carry out the sums
over the group elements (note that given an edge $e$, the groups elements $g_e$ and $g_e^{-1}$
appear in $\prod_{f} w_f$ as many times as number of faces share the
edge $e$). We obtain the following expression for the partition function
\be\label{Z-SF}
 Z=\sum_{\{\rho_f\}}\big(\prod_{f}\tilde w_{\rho_f}\big)\!\!\!\!\!\!\!\!
\sum_{\,\quad\{a^f_v=1\}{}_{v\subset f}}^{\text{dim}\rho_f}
\prod_e
\tilde{P}^e_{a^{f^+}_{s},...,a^{f^-}_{s},...;a^{f^+}_{t},...,a^{f^-}_{t},...}(\{\rho_f\}
_{f\supset e}),
\ee
where, associated to every edge, we have defined the projector 
\be\label{int}
 \tilde{P}^e_{a^{f^+}_{s},...,a^{f^-}_{s},...;a^{f^+}_{t},...,a^{f^-}_{t},...}
(\{\rho_f\}_{f\supset e}):=\frac1{|G|}\sum_{g_e}\prod_{{{f^+}\supset
 e}}\rho_{f^+}(g_e)_{a^{f^+}_{s}a^{f^+}_{t}}
 \prod_{{{f^-}\supset e}}\rho^*_{f^-}(g_e)_{a^{f^-}_{t}a^{f^-}_{s}}.
 \ee
In the above tensor, the first and third groups of indices (distinguished by the superindex $f^+$)
involve all the faces that have $e$ in their boundary with the same orientation as the face.
These groups of indices have therefore as many indices as number of faces with $\varepsilon^f_e=1$.
In turn,
the second and fourth groups of indices (distinguished by the superindex $f^-$) have as many indices
as number of faces $f$ with $\varepsilon^{f}_e=-1$. We show an example in figure
\ref{fig:curvature}b. To every pair face--edge two indices are
associated, $a^f_{s}$ and $a^f_{t}$, that we  attach to the vertices of $f$
that bound the edge $e$, namely $s(e)$ and $t(e)$.
In equation \eqref{Z-SF}, for every face $f$ of the two-complex, there is a sum for every vertex $v$
belonging to that face. Note that every index $a^f_{v}$ appears twice, since there are two edges
meeting at the vertex $v$ and bounding the face $f$. Then, the product over edge projectors contracts
all the indices.

Associated to every edge there is a Hilbert space $\mathcal{H}_e$.
Let us denote by $f_1,\cdots f_m$ the faces attached to $e$. Then,
\be
\mathcal{H}_e\;:=\;V_{\rho_{f_1}}\otimes V_{\rho_{f_2}}\otimes\ldots\otimes V_{\rho_{f_m}}.
\ee
Here, to keep the notation simple, we are assuming that $\varepsilon^{f}_e=1$ for all the
faces\footnote{If there would be some face with opposite orientation to that of $e$, we would
replace the corresponding vector space $V_{\rho_f}$ for its dual $V^*_{\rho_f}$.}.
The corresponding tensor $\tilde{P}^e$ defines an orthogonal projector onto the gauge--invariant
subspace of the edge-Hilbert space $\mathcal{H}_e$
\be
(\tilde{P}^e\psi)_{a_1a_2\cdots a_m}:=(\tilde{P}^e)_{a_1a_2\cdots a_m;\;b_1b_2\cdots
b_m}\psi_{b_1b_2\cdots b_m}
\ee
for $\psi \in \mathcal{H}_e$.  Here, as for the spin net models, the projector (\ref{int}) is the Haar-intertwiner. As before, more general spin foam models are constructed by restricting the Haar intertwiner to proper subspaces of the gauge invariant subspace of $\mathcal{H}_e$.

For spin foam models one usually reorganizes the partition function (\ref{Z-SF}) such that
amplitudes can be associated to vertices. To this end, one decomposed the projectors $\tilde P^e$ by
 introducing an orthonormal basis $\iota^{e_k}$, $k=1,\ldots,m$, for the invariant subspace of
$\mathcal{H}_e$. By adjusting the basis we can decompose any gauge invariant projector  as (here $m'\leq m$ and $m'=m$ for the Haar intertwiner)
\be\label{Gl:Decomposition}
\tilde{P}^e\;=\;\sum_{k=1}^{m'}\;|\iota^{e_k}\rangle\langle\iota^{e_k}|,\qquad
 \tilde{P}^e_{a_1a_2\cdots a_{m'};\;b_1b_2\cdots
b_{m'}}=\sum_{k=1}^{m'} \iota^{e_k}_{a_1a_2\cdots a_{m'}}\iota^{e_k}_{b_1b_2\cdots b_{m'}},
\ee
so that the indices attached to $t(e)$ are now carried by the intertwiners $|\iota^e\rangle$ and the
indices attached to $s(e)$ are carried by the intertwiners $\langle\iota^e|$.
According to the description above, we can contract within every vertex the corresponding
$\iota^e$ obtaining vertex amlitudes $\mathcal{A}_v(\rho_f,\iota^e)$, that depend on the
representations $\rho_f$ and intertwiners $\iota^e$ associated to the faces and edges that meet at
$v$. 
With this, the partition function can be written in terms of vertex amplitudes via
\be\label{Z-SF-vertex}
Z\;=\;\sum_{\rho_f\,\iota^e}\;\prod_{f}\tilde w_{\rho_f}\;\prod_v\mathcal{A}_v(\rho_f,\iota^e).
\ee
This is the usual description employed in spin foam models for quantum gravity.

As for the spin net models, we can also define a holonomy representation 
\cite{Pfeiffer:2001ig,Pfeiffer:2001ny, bahrta} for spin foams using the inverse group Fourier
transform. In case we are dealing with a non--trivial edge projector
(\ref{Gl:Decomposition}), i.e.\ not the Haar intertwiner, the resulting holonomy representation will
be of the form of a generalized lattice gauge theory. That is, instead of only one group variable
associated to every edge we will have as many group variables attached to a given edge as there are
faces attached to this edge. Furthermore, there will not only be face weights $w_f$ but also edge
weights $P^e$ which result from the transformed edge projectors $\tilde P^e$.

Let us now consider the case of Abelian groups $\Zl_q$. As for spin net models, the spin foam
representation simplifies since the irreducible representations are just one--dimensional and the
group Fourier transform is just given by the usual discrete Fourier transform.  Namely, we obtain

 \ba\label{y4}
Z
&=&   \sum_{\{k_f\}}  \prod_f \tilde w_{k_f}   \prod_e  \,  \delta^{(q)} \!\big(
\sum_{f\supset e} \varepsilon^f_ek_f\big) \nn\\
&=&   \sum_{\{k_f\}}   \prod_f \tilde w_{k_f}  \prod_v \prod_{e\supset v} 
\delta^{(q)} \!\big( \sum_{f\supset e} \varepsilon^f_e k_f\big),
\ea
where the $q-$periodic delta function $ \delta^{(q)}(\cdot)$ enforces the
Gau\ss~constraints, now based on the edges (instead on the vertices as for spin nets).
In the last step we just splitted the delta functions over every edge into
two delta functions over every vertex, in order to define the vertex amplitudes
(which here are $ A_v=\prod_{e\supset v}  \delta^{(q)} \!\big( \sum_{f\supset e}
\varepsilon^f_ek_f\big)$).

In case the face weights are given by $\tilde w_{k_f}=1$, which is the $BF$ theory case, we
obtain an additional symmetry for the partition function (\ref{y4}), which as commented
before is known as translation symmetry. For spin foams the gauge parameters $k_c \in \Zl_q$ are
based on the 3--cells of the lattice (here we assume that the 3--cells are well defined, for a
regular hypercubic lattice these would be the 3D cubes):
\ba\label{transisf}
k_f \mapsto k'_f=k_f + \sum_{c \supset f} \varepsilon^c_f k_c
\ea
where $\varepsilon^c_f=+1$ ($=-1$) if the orientations of the 3--cell $c$ agrees (disagrees) with
the one of the 2--cell $f$. As for the spin net models this translation symmetry does not change the
validity of the Gau\ss~constraints appearing in the partition function (\ref{y4}).  For the
gravitational spin foam models (in 3D) this type of gauge symmetry gives the diffeomorphism symmetry
underlying general relativiy \cite{Freidel:2002dw,Dittrich:2008pw}. It hence has a special status and is
indeed deeply intertwined with triangulation or more generally discretization independence of the
models \cite{Dittrich:2008pw, Bahr:2009ku, Bahr:2009mc, Bahr:2009qc,Bahr:2011uj}.

For coarse graining we will consider the Abelian cutoff models with face weights coinciding with the
edge weights in spin net models (\ref{Aco}). For the cutoff models the translation symmetry will be
broken as the weights are now no longer constant in $k_f$. The motivating question will be to see
whether these models flow back to the $BF$ phase, in which the translation symmetry is restored. 

\begin{figure}
\begin{center}
\includegraphics[width=7cm]{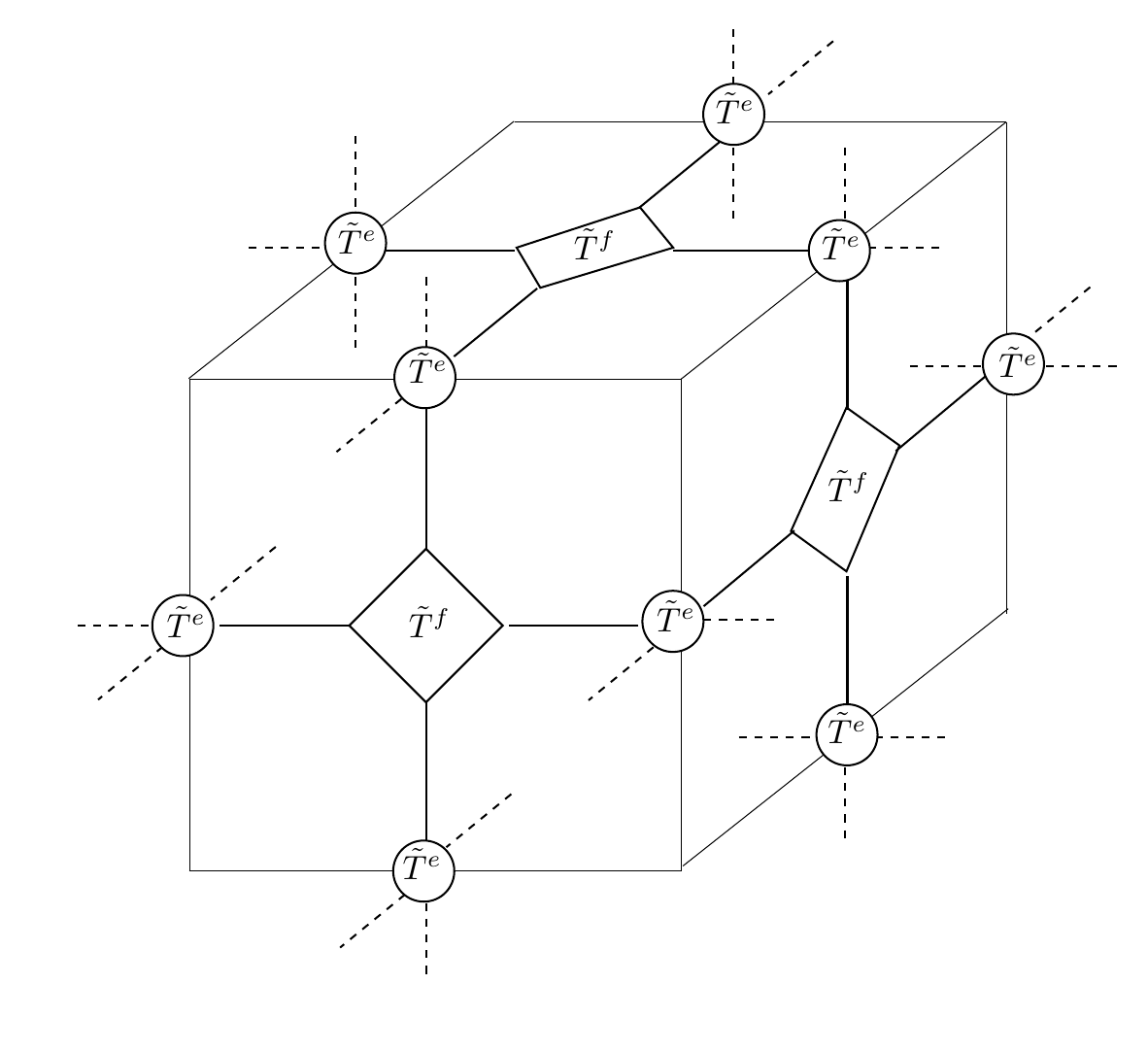}
\caption{Schematic definition of the tensor network for spin foams on a
cubic lattice.
\label{tnw-sf}}
\end{center}
\end{figure}

Finally let us note that also spin foams can be represented in various ways as tensor networks. One
possibility would be to start with the representation \eqref{Z-SF-vertex} involving vertex
amplitudes. To obtain an algebraically similar form to the tensor network representation of spin net
models one would however keep the edge projectors $\tilde P^e$ intact. To this end we associate the
tensor $\tilde T^e=\tilde P^e$  to the midpoints of the edges of the lattice. The
edges of the lattice carry a number of indices which are all contracted with each other in the
lattice vertices according to the description below (\ref{int}). Furthermore we have to take care of
the face weights $\tilde w_{\rho_f}$ and the sum over representation labels $\rho$. This can be
achieved by introducing another type of tensor $\tilde T^f= \tilde w$, see figure
\ref{tnw-sf}. If we work with a
cubic lattice these tensors are four-valent carrying as indices representation labels:  $(\tilde
T^f)_{\rho_1\rho_2\rho_3\rho_4}= \tilde w_{\rho_1} \delta(\rho_1,\rho_2,\rho_3,\rho_4)$.  The
second factor is equal to one if all representation labels in the argument coincide and zero
otherwise. This tensor is connected by auxiliary edges to the four adjacent edge tensors $T^e$
ensuring that the sum over the representation labels involves the same representation label for
every face. 

\subsection{Relation between spin foam and spin net models}
\label{subsec:relation}

Here we want to comment on an interesting relationship between spin net and spin foam models.
Namely, one can understand spin nets as expectation values of (gauge symmetry breaking) observables
inserted into the spin foam partition functions. We will illustrate this only for the simplest case:
the spin foam represents $BF$ theory and the spin net is of the form (\ref{Z-hol1}), i.e.\ the
vertex projector is given by the Haar intertwiner.

 Indeed, the partition function
\eqref{Z-hol1} can be rewritten in the form
\begin{align}\label{observable}
Z&=\frac1{|G|^{\sharp e}}\sum_{\{g_e\}} \prod_e w_e(g_e)
\frac1{|G|^{\sharp v}}\sum_{\{g_v\}}\prod_e \delta_G(g_e^{-1}g_{s(e)}g_{t(e)}^{-1})\nonumber\\
&\propto\frac1{|G|^{\sharp e}}\sum_{\{g_e\}} \prod_e w_e(g_e) \prod_{f} \delta_G(h_f),
\end{align}
where $f$ runs over a set of faces whose boundary edges and vertices generate the initial
graph over which the spin net model is defined. Moreover, the two-complex made up by this set
of faces must be simply connected for equation  \eqref{observable} to be valid (otherwise we have to amend the condition that holonomies along non--contractible loops should be trivial). The constant of
propotionality comes from the
normalization of the delta functions and can be absorbed by a redefinition of the weights $w_e(g_e)$
in the second line.
The second line of equation  \eqref{observable} is the result of introducing the product of edge weights 
$w_e(g_e)$ in the partition function for  $BF$ theory
($w_f=\delta_G$) on a simply connected two-complex.
Similar observables have been considered in the context
of
3D quantum gravity, namely for the Ponzano-Regge model \cite{Barrett:1998dc,Barrett:2011qe} and have been interpreted as Feynman diagram evaluations.
That is, the edge-weights $w_e(g_e)$ can
be understood as propagators and the spin net model as a Feynman diagram evaluation.

\section{Coarse graining methods} \label{coarsemethods}

Having introduced the models of interest in the previous section, we now turn to the challenge of implementing a coarse graining procedure. In this section, we will outline our general approach and discuss the conceptual issues that arise. The application of two approximation schemes in our context will then be discussed in the following sections.

Although gravitational spin foams are usually defined on a general triangulation or two--complex we will here consider coarse graining on a regular lattice. This allows us to actually make explicit computations and to use methods from lattice gauge theory and condensed matter systems. One might object that using a regular lattice will introduce a background structure, spoiling background independence. There are, however, indications \cite{Bahr:2009qc,Bahr:2011uj,Rovelli:2011fk} that a restoration of diffeomorphism symmetry will be connected with a notion of triangulation or discretization independence, and hence the choice of a particular underlying lattice may not matter.
Indeed, the universality phenomenon of statistical systems also suggests that the details of the chosen lattice might not matter for the questions we are interested in, i.e.\ a characterization of the possible phases of the models, or whether spin foams can avoid degenerate phases. Nevertheless one should study whether the results depend on the choice of lattice. 

The development of a scheme where order parameters or coupling strengths might be locally varying is another conceptual challenge \cite{Markopoulou:2000yy,Markopoulou:2002ja}, which we will not address here. This would be appropriate for a random lattice or for situations with a very inhomogeneous dynamics. Again, we think that developing feasible coarse graining methods for spin foams and nets on a regular lattice is an indispensable first step.

Furthermore, for gravitational systems, a regular underlying lattice can nevertheless represent very inhomogeneous or irregular geometries. This is due to the fact that the dynamical variables are the geometrical data which also determine the lattice geometry and moreover the physical scale. In this sense a very fine lattice can carry very different (coarser) geometries \cite{Dittrich:2008pw, Rovelli:2010qx}.\footnote{See also the universality result \cite{DelasCuevas:2009gc}, which  implies the simulation of irregular  sublattices based on a regular underlying lattice.}  This opens up the possibility that refining in lattice quantum gravity is even equivalent to summing over (a class of coarser) lattices \cite{Rovelli:2010qx}, thus eventually leading to a discretization independent theory at the fixed point \cite{Bahr:2009qc,Bahr:2011uj,Rovelli:2011fk}. Indeed, systems with diffeomorphism (like) symmetry might add interesting new insights to the theory of phase transitions.

Another issue, which has to be addressed for any coarse graining or renormalization scheme, is which
space of models one considers the renormalization flow to take place in \cite{Oeckl:2002ia}. Indeed,
this question is not quite obvious for spin foam or spin net models as, for instance, spin net
models mix aspects of edge models, where couplings are along edges,  with vertex models, where
couplings are based at the vertices.
type of degrees of freedoms couple in a certain way might not be available for the non--trivial
models.

Furthermore, spin foam models are constructed to be as independent as possible from the underlying
discretization. This translates into certain invariance properties of the amplitudes under a certain
class of subdivisions, those which effect edges or plaquettes, but do not lead to an increase in the
number of (non--trivial) vertices \cite{Bojowald:2009im,Bahr:2010my,Bahr:2010bs,Conrady:2005qu}.
Spin foam models, which are constructed using trivial face weights but non-trivial projectors,
will satisfy this invariance. Invariance under this subclass of subdivisions can be seen as a first
step towards a discretization independent model.
One might ask
whether it is possible to come up with a renormalization scheme in which this form of the spin foam
amplitudes and the invariance properties under face and edge subdivisions is preserved, see also
\cite{bahrta}.  However,  this seems not to be very likely, as long as the amplitudes are not
invariant under all kinds of subdivisions and hence renormalization flow is trivial or at a fixed
point. An intuitive reason is that the faces and edges of a coarse grained spin foam are `effective'
building blocks, containing a huge number of bare vertices, edges or plaquettes.
That is, a change of the effective triangulation, even if it only involves a subdivision of edges and faces, will correspond to a more complicated change of the underlying `bare' triangulation. Indeed, we will find that under the renormalization schemes presented here, the invariance property under edge and face subdivisions is not preserved.

A general problem with real space renormalization schemes of (higher than one--dimensional)
statistical models is that an increasing number of non--local couplings appear with each blocking
step.\footnote{Indeed, such non--local couplings are essential to regain diffeomorphism
\cite{Bahr:2010cq} or Lorentz symmetry \cite{Hasenfratz:1993sp,dowker}. This at least applies to
models with genuine field degrees of freedom, i.e.\ 4D gravity, but not for 3D gravity, which is a
so--called topological field theory. Nevertheless, 3D gravity (which can be described by $BF$
theory, that is  the low temperature fixed point of lattice gauge theories) is an important test
case for which the restoration of symmetries can be studied in a simpler setting.}
To keep the renormalization flow in a space with a finite number of parameters, some sort of truncation or approximations scheme has to be used. We will consider two such schemes. The first one, the Migdal-Kadanoff \cite{Migdal:1975zg,Kadanoff:1976jb} scheme, is based on a truncation to local couplings. The derivation of this scheme is based on arguments that rely on the standard form of edge and plaquette models. Here an important question for future work would be to generalize this scheme to proper spin foam or spin net models, which constitute a rather generalized form of these models.

Whereas the Migdal-Kadanoff scheme is based on a blocking of the group variables, the second scheme,
based on tensor network renormalization, is blocking the degrees of freedom encoded in the
representation labels. This might be an advantage as the representation labels are related to metric
degrees of freedom in the geometric spin foam models. Hence, this kind of  blocking is much closer
to the  blocking used in the gravitation models in \cite{Bahr:2009qc,Bahr:2010cq}, which is derived
by geometrical arguments. Moreover we will see that this scheme allows direct access to the behavior
of the vertex and edge projectors and makes use of the properties of representation theory
encoded in the models.

The tensor network renormalization can, however, also be applied in the holonomy representation of
the spin foam and spin net models, see also \cite{vidalgauge}. Which scheme to use might depend on
which region one aims to explore, the high temperature (where the spin net/ spin foam representation
is more appropriate) or the low temperature region (where the holonomy representation might be more
useful). Indeed, tensor networks are an extremely general tool, in which it is also possible to
consider the emergence of different kinds of effective degrees of freedom, long range order as well
as topological phases \cite{wenemer}. Another advantage is that tensor networks can handle
non--positive weights \cite{vidalgauge}. This is an important point for gravitational spin foams
where oscillating amplitudes appear, preventing the use of Monte Carlo simulations.

The issue of non--local couplings in this renormalization scheme appears in the form of having more and more degrees of freedom. One has to choose a cutoff for this number. The advantage as compared to the Midgdal Kadanoff scheme is that the accuracy of this scheme can be systematically improved by increasing this cutoff. Furthermore, it is possible to study the dependence of the results on the choice of cutoff.

On the other hand the tensor network renormalization scheme requires much more effort than the Migdal-Kadanoff scheme and offers less analytical control. This is of course related to the much bigger parameter space one is considering in the case of tensor networks. A crucial question for future research is how feasible the tensor network scheme will be for higher dimensional systems, as most work  performed so far is for one and two--dimensional systems. Our application of the tensor network renormalization method is also restricted to two--dimensional spin net models.


\section{The Migdal--Kadanoff approximation}\label{mk}

Migdal-Kadanoff (MK) approximations are a simple tool to overcome a key difficulty in real-space renormalization: the introduction of non-local couplings in the renormalized action (which in the statistical physics language is the Hamiltonian) with each coarse graining step.

When coarse graining the one-dimensional Ising model with nearest-neighbour interactions by decimation, the state sum factorizes. Integrating out the chosen spins gives a renormalized Hamiltonian of the same form as the one we had started with \cite{Kadanoff:1976jb}. However, this no longer happens in higher dimensions: generically, the renormalized Hamiltonians feature new, longer-ranged interaction terms that render it impossible to continue with the procedure \cite{Migdal:1975zg}. The central idea behind the MK approximation is to substitute this renormalized Hamiltonian by an approximate Hamiltonian which features the same interaction terms as the original Hamiltonian, thus making the renormalization transformation \emph{form-invariant}.

\begin{figure}
\begin{center}
 \mbox{
      \subfigure[isotropic decimation]{\includegraphics[width=0.35\textwidth]{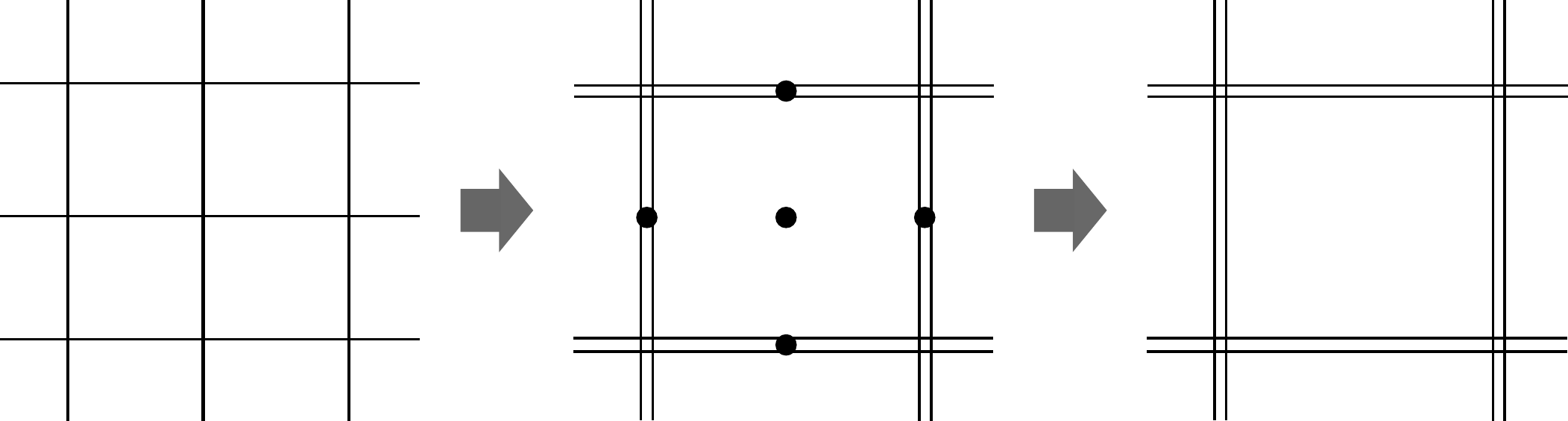}} \quad
      \subfigure[anisotropic decimation]{\includegraphics[width=0.60\textwidth]{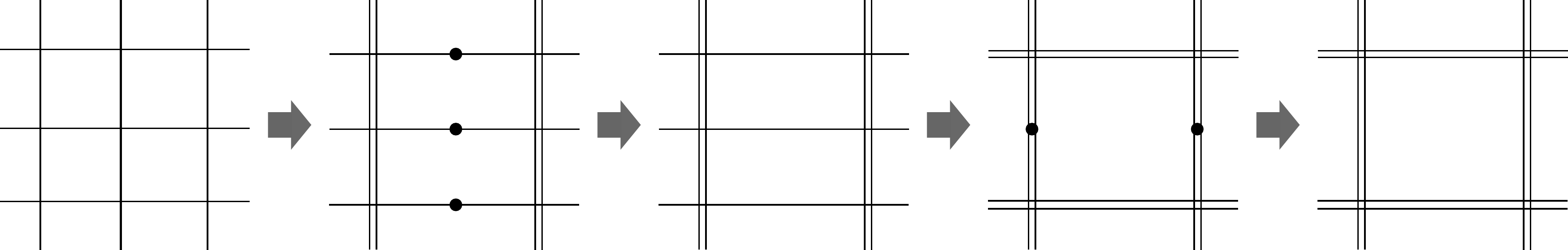}}
      }
\caption{Derivations of MK relations for 2d spin net models involve moving bonds and subsequent integrations. The isotropic procedure leads to exponents $\gamma = 1, \lambda =2$, whereas in the anisotropic case this is only true for the couplings in the vertical direction -- the horizontal couplings feature $\gamma = 2, \lambda=1$.  \label{fig:mk}}
\end{center}
\end{figure}

The original proposal \cite{Migdal:1975zg} of  Migdal amounts to neglecting certain terms in the state sum, while strengthening others. Applied to the two-dimensional Ising model, this translates to moving bonds (couplings along the edges) away from those nodes which are to be eliminated by decimation. In this way the valency of those particular nodes is reduced to two and the situation is thereby effectively reduced to the one-dimensional case (see figure \ref{fig:mk}). 

Kadanoff subsequently identified bond-moving as a special case of a general approximation scheme for
which he derived a bound on the free energy  \cite{Kadanoff:1976jb}. Further justification for this
approximation is derived from Monte Carlo simulations \cite{Creutz:1979kf,Creutz:1979zg}. In
practice, MK approximations do fairly well at finding fixed points and phase transitions and not so
well at determining the order of these transitions \cite{Creutz:1982vz}. However, specific fixed
points of Kosterlitz-Thouless type might not be reproduced by this approximation \cite{Ito:1985bv}.
MK approximations are computationally efficient and comparably easy to implement but not refineable
in a systematic way.

\subsection{Migdal-Kadanoff relations explored} \label{mkex}

Let as consider a model of spin net type
\ba \label{mk-simple-model}
  Z \sim \sum_{\{g_e\}} \prod_e w(g_e)
\ea
based on the Abelian group $\Zl_q$. Bond--moving then involves dropping some of the terms in
(\ref{mk-simple-model}) and replacing some of the others with
\ba
  w(g_{e'}) \rightarrow w^2(g_{e'}) = \sum_k  \left(  \sum_j \tilde{w}(k-j) \tilde{w}(j) \right) \chi_k(g_{e'})
\ea
leading to convolution of Fourier coefficients, the main feature of MK approximations.  The subset of group variables corresponding to dropped terms can then be integrated out without generating non--local couplings. 

One can proceed similarly for lattice gauge theories or plaquette models of the form (\ref{Z-SF-hol1}). For both kind of models different versions of bond--moving and different decimation schemes can be defined. In particular, in anisotropic methods the renormalized couplings will be different in different lattice directions, while in simpler isotropic methods the couplings will remain homogeneous (see figure \ref{fig:mk}). We will only consider the isotropic methods here, which can also be made exact on so--called hierarchical lattices \cite{Ito:1984ph}.

For the normalized weights $Q(k) := \tilde{w}(k)/\tilde{w}(0)$ this results in general recursion relations of the form \cite{Ito:1985bv}
\ba \label{mk-rr}
 Q_{n+1}(k) = \left( \frac{\sum_{j = 0}^{q-1} Q_{n}^\gamma(k-j) Q_{n}^\gamma(j)}{\sum_{j = 0}^{q-1} Q_{n}^\gamma(-j) Q_{n}^\gamma(j)} \right)^\lambda .
\ea
Here, $\gamma$ and $\lambda$ are constants specific to the model and derivation in question. However, models with fixed $\gamma \cdot \lambda$ are qualitatively equivalent and share the same fixed point structure in the following sense: Two instances of (\ref{mk-rr}), with exponents $\gamma,\lambda; \tilde{\gamma},\tilde{\lambda}$ resp.\ and $\gamma \cdot \lambda = \tilde{\gamma} \cdot \tilde{\lambda} $ are related by
\ba
  Q_{n}(k) = [\tilde{Q}_{n}(k)]^{\lambda/\tilde{\lambda}}
\ea
given initial configurations that satisfy
\ba
 \tilde{Q}_0(k) = [Q_{0}(k)]^{\gamma/\tilde{\gamma}}.
\ea

Typical values for the exponents are $\gamma = 1$ and $\lambda = 2$ for `isotropic'  2D spin net and 4D lattice gauge models and $\gamma = 1$ and $\lambda = 4$ `isotropic' 3d lattice gauge theories. The corresponding fixed point equations are given by
\ba \label{fp-eq}
 Q(k) \left( \sum_{j = 0}^{q-1} Q^\gamma(-j) Q^\gamma(j) \right)^\lambda - \left( \sum_{j = 0}^{q-1} Q^\gamma(k-j) Q^\gamma(j) \right)^\lambda = 0 .
\ea

The high and low temperature fixed points ( $Q(0) = 1, Q(k>0) = 0$ and $Q(k) = 1 \, \forall k$, resp.) are common solutions of eq.\ (\ref{fp-eq}), independently of the exponents. Furthermore, for some fixed factor $d$ of $q$, there is a class of fixed points given by 
\ba
 Q(k) = \begin{cases} 1 \qquad k \mbox{ mod } d = 0 \\
                      0 \qquad \mbox{else }
        \end{cases}
\ea
which derive this property from being invariant under $d$-fold cyclic permutations.

\begin{figure}
\begin{center}
 \mbox{
      \subfigure[$\lambda=2$, spin net 2D]{\includegraphics[width=0.45\textwidth]{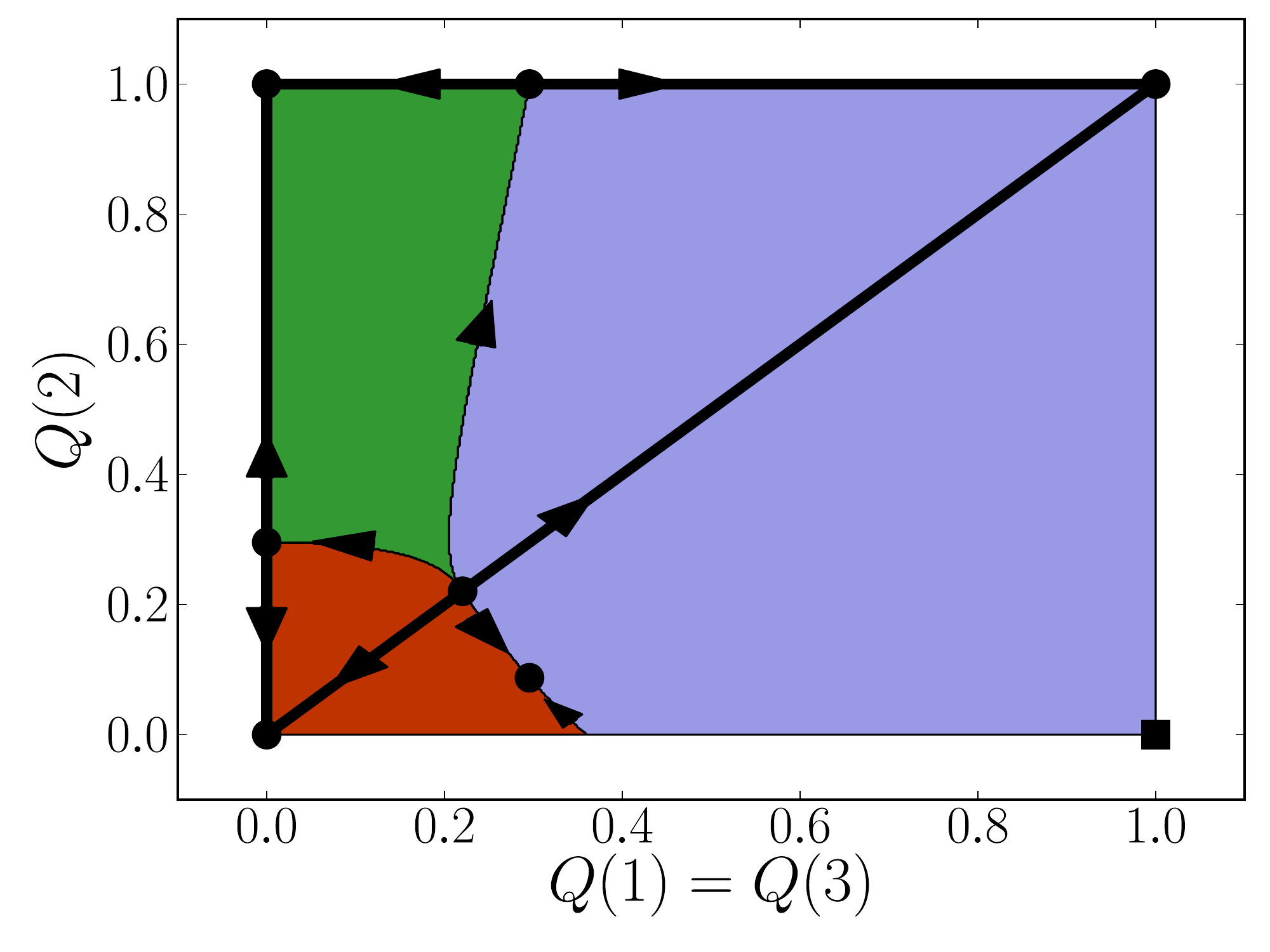}} \quad
      \subfigure[$\lambda=4$, spin foam 3D]{\includegraphics[width=0.45\textwidth]{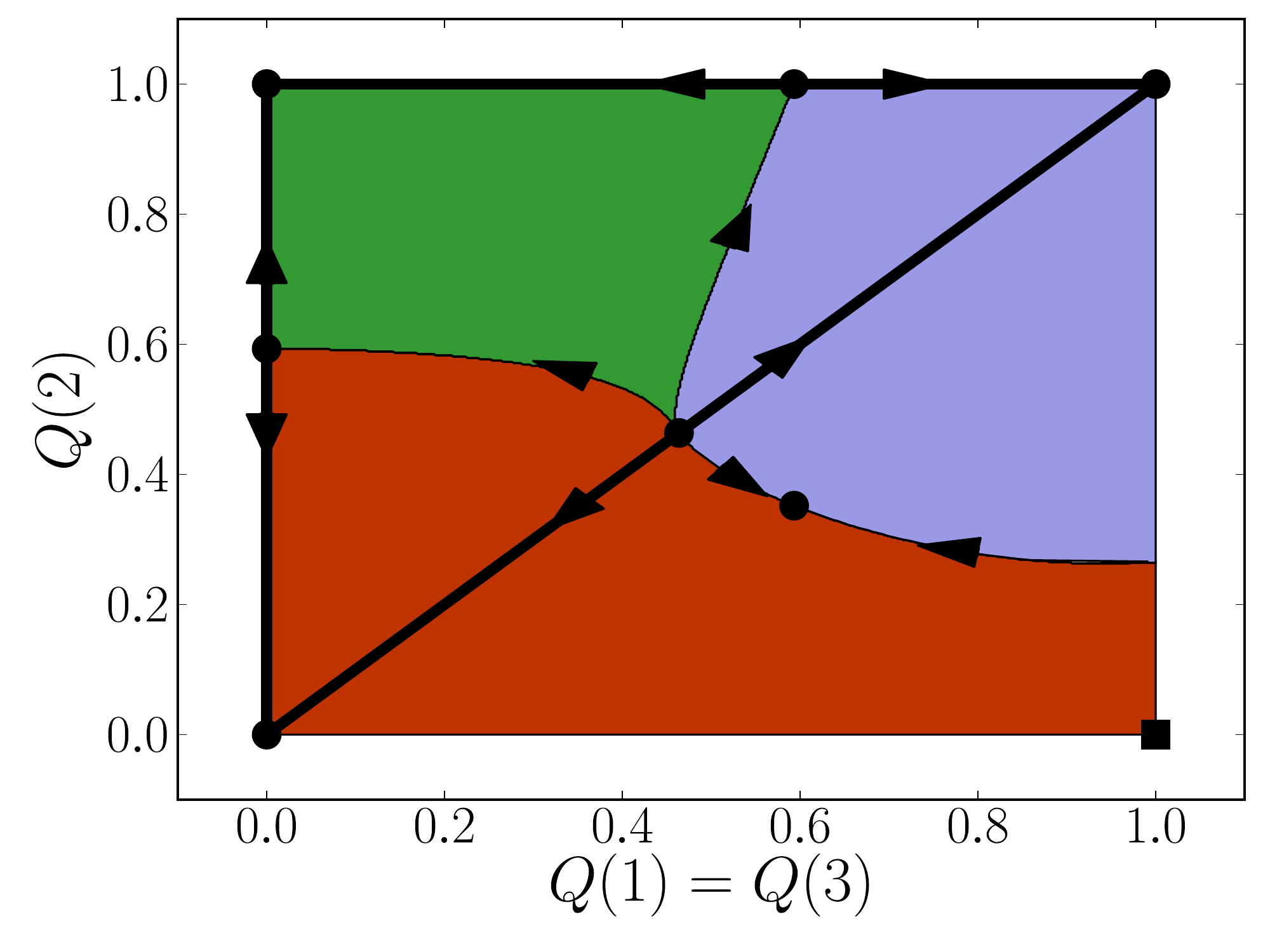}}
     }
\caption{Parameter space of $\mathbb{Z}_4$ symmetric models and their renormalization flow. The three regions correspond to configurations flowing to HTF $(0,0)$, LTF / BF $(1,1)$ and the `cyclic fixed point' $(0,1)$. Other unstable fixed points (dots) and invariant submanifolds (thick lines) are also depicted. Note the differing behaviour of the cutoff-model (square) at $(1,0)$ in both cases. \label{fig:mk-z4}}
\end{center}
\end{figure}

More generally, for a fixed factor $d$ of $q$, (symmetric and normalized) 
configurations with $Q(k) = 0$ for $k \mbox{ mod } d \neq 0$ and those given by 
\ba
Q(k) = \begin{cases} 1 \qquad k \mbox{ mod } d = 0  \\
                     \alpha \qquad \mbox{else }
       \end{cases}
\ea
form 
invariant submanifolds of the parameter space. There are also more nontrivial fixed points, including the higher dimensional analogue of the Ising fixed point on the one-dimensional invariant line connecting the low and high temperature fixed points. For $\mathbb{Z}_2$, this point is predicted to be a nontrivial solution of 
\ba
  Q (1+ Q^{2\gamma} ) ^\lambda - (2 Q^\gamma)^\lambda = 0.
\ea 
In the 2D standard Ising model case (with exponents $\gamma=1,\lambda=2$) the solution is given by $ Q = \tilde{w}_1 / \tilde{w}_0 = 0.296$ which corresponds via $ kT = \mbox{Artanh}(Q)^{-1}$ to a temperature of $ kT = 3.282 $ (the exact solution is given by $ kT_c = 2/\log(1 + \sqrt{2}) \approx 2.269$ \cite{Onsager}). Also note that MK approximations only predict this Ising-type fixed point for exponents with $\gamma \cdot \lambda > 1$. Figure \ref{fig:mk-z4} illustrates the afore-mentioned features for 2D spin net and 3D spin foam models with group $\mathbb{Z}_4$. 

\subsection{Restoration of $BF$ symmetry in Abelian cutoff models}\label{mkaco}

Cutoff models are certain initial configurations for the MK renormalization group flow which are of
particular interest to quantum gravity because they are derived from $BF$ theory. 
The question of interest here is whether the topological nature of $BF$ theory, which is destroyed
in the construction of the cutoff models, will be restored under the renormalization group flow.
With a possible restoration of the $BF$ phase, also the translation symmetries
(\ref{transisn})-(\ref{transisf}) would be restored, which in the 3D gravity models correspond to
diffeomorphism symmetry. Whereas the $BF$ phase ---or low temperature fixed point (LTF)--- in a 3D
gravity setting represents flat space time, the high temperature fixed point (HTF), at which $Q(0)$
is equal to unity and vanishes for all other labels $k$, corresponds to a geometrically degenerate
phase, in which all geometric (length) observables have vanishing expectation value.
However, whether the models flow back to $BF$ or not depends very much on the initial configuration in question. Instead of focussing on one model with one specific group, we here address the differences between different models for finite Abelian groups $\mathbb{Z}_q$ with varying $q$, leaving other (non-Abelian) groups for further research.

\begin{figure}
\begin{center}
\mbox{
     \subfigure[spin net, $\lambda=2$]{\includegraphics[width=0.50\textwidth]{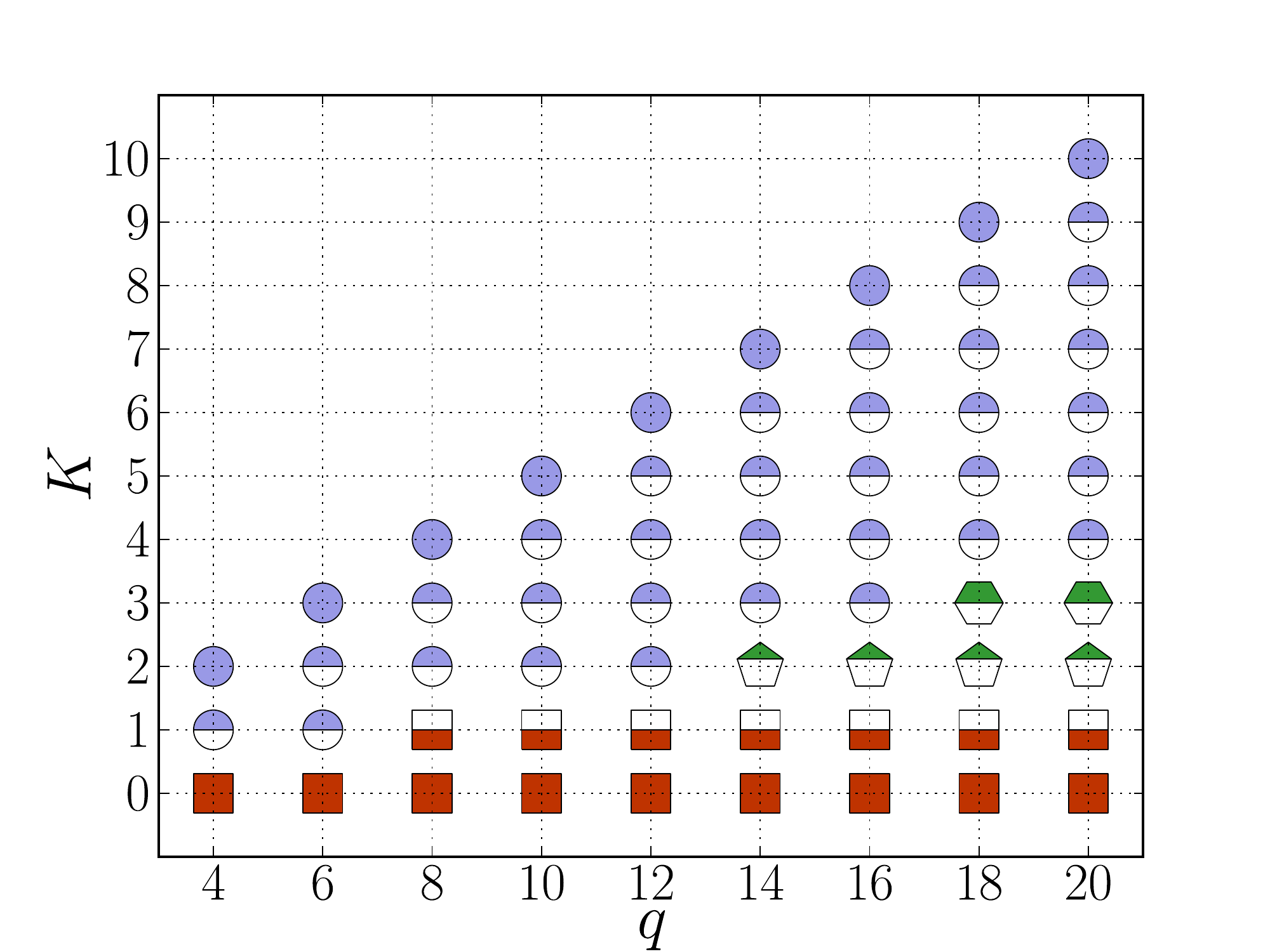}} \quad
     \subfigure[spin foam, $\lambda=4$]{\includegraphics[width=0.50\textwidth]{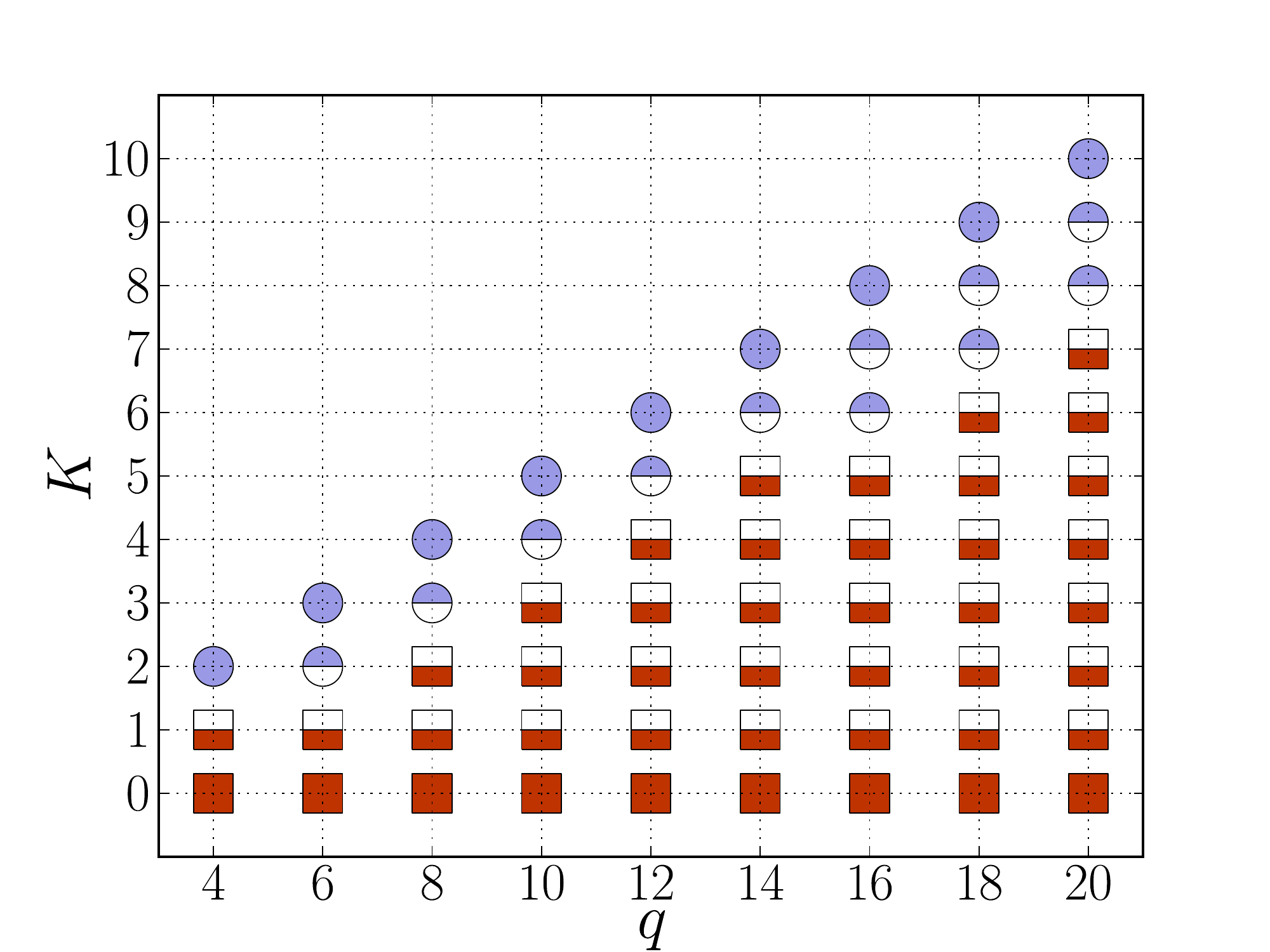}}
     }
\caption{Flow behaviour of different cutoff models, labelled by $K$ and $q$. Markers half-filled at the top flow to BF/LTF (diagonal models, $K = q/2$), markers half-filled at the bottom flow to HTF (horizontal modesl with $K = 0$). In (a), unstable fixed points appear (pentagonal and hexagonal markers).\label{fig:mk-Kq}}
\end{center}
\end{figure}

Let us first consider the case of 3D lattice gauge theory / spin foam models over $\mathbb{Z}_q$ for varying $q$ where the exponents in (\ref{mk-rr}) are given by $\gamma = 1, \lambda = 4$. Here, different initial configurations parametrized by $q$ and $K$ converge quite fast (after 5 to 10 iterations) either to the HTF or to LTF, see figure \ref{fig:mk-Kq}b. In general, there are two competing effects encoded in the recursion relations (\ref{mk-rr}): the convolution leads to a broadening of the function $Q(k)$, whereas the exponents $\gamma$ and $\lambda$ lead to a dampening effect (as the $Q(k)\leq 1$). 

For the 3D gauge models most configurations flow to the HTF, that is the dampening factor is quite
strong. Regarding the question of restoration of the translation symmetries we see that this happens
only in cases where the initial configuration is already quite close to the $BF$ configuration,
which
coincides with the LTF. 

These observations are in accordance with similar, but analytical work done in the generalized case of $U(1)$ \cite{Ito:1985bv}. There it is found that for $3D$  $U(1)$ gauge models all configurations satisfying certain conditions\footnote{ These include a  positivity requirement on the weight functions both in representation space and group space which is not satisfied for our cutoff models. However it would be satisfied for instance for a heat kernel regularization of for instance the Ponzano--Regge model. This would show that with the Migdal-Kadanoff method these regulated models would not flow back to the Ponzano--Regge model, but to the high temperature fixed point.} flow to the HTF. 

These results have been extended to 3D lattice gauge models with non--Abelian compact groups $U(N)$
and $SU(N)$ by \cite{Muller:1984uz}. Hence this applies also to the 3D gravitational
(Ponzano--Regge) model, which is based on $SU(2)$ (assuming a change from a lattice based on
tetrahedra to a cubic lattice does not matter) and we have to conclude that within the
Migdal-Kadanoff approximation translation symmetry / diffeomorphism symmetry is not restored. Note,
however, that for finite groups (i.e.\ finite $q$), there are some (even) cutoff configurations that
do flow back to the LTF or $BF$ phase. Here we might draw the conclusion that it is easier to
restore a compact  symmetry as compared to a non--compact one, as the translation symmetry is based
on a compact parameter space for $\mathbb{Z}_q$ and on a non--compact one for proper Lie groups.
Here it would be interesting to see whether for some analogous modifications of the Tuarev--Viro
models \cite{Turaev:1992hq}, which describe 3D 	quantum gravity with a cosmological constant and are
based on quantum groups, such a restoration of the (here compact) translation symmetries occurs.  
See also a discussion of related issues in loop quantum gravity quantization of 3D gravity with a cosmological constant \cite{Perez:2010pm,Noui:2011im} and a classical coarse graining treatment of the same system, where diffeomorphsim symmetry can be restored \cite{Bahr:2009qc}. 
To study this question for the Tuarev--Viro models one would have to adjust the Migdal-Kadanoff method to quantum groups or to apply alternative renormalization schemes, such as the tensor network scheme described in the next section.

For 4D  `isotropic' lattice gauge models, or equivalently for 2D spin net or edge models, the exponents in \eqref{mk-rr} are given by $\gamma = 1, \lambda = 2$. Hence the dampening effect is much less pronounced than for the 3D gauge models. Indeed, now most configurations flow back to the LTF or $BF$ phase, see figure  \ref{fig:mk-Kq}a. 
 An exception are the $K=1,q \geq8$ configurations which flow to HTF, however only after a considerably large number of iterations (around 60). Nevertheless this can be interpreted as a phase boundary in the $q-K$ diagram. 

For a given cutoff $K>1$ the number of iterations necessary to converge to the LTF grows with $q$.
(For $K=2$ these are $8,15,60$ iterations for $q=8,10,12$ respectively). Moreover, for sufficiently
large $q$, the simulations go through a long phase of only very small changes of the order
$10^{-5}-10^{-4}$, so that the iterations are almost constant. For $K=2$ this appears starting with
$q\geq 14$. For instance from around iteration 10 to iteration 100 the following configurations
appear for the $q=14$ simulation
\ba\label{pex1}
\textstyle  Q = {\scriptstyle (1,\, 0.81,\,0.43,\,0.15,\,0.03,\,0.005,\,0.0005/6,\,0.0001,\,0.0005/6,\,0.005,\,0.03,\,0.15,\,0.43,\,0.81) }
\ea
and would be stable at least up to the number of digits displayed in (\ref{pex1}). This configuration converges after 970 iterations to the low temperature fixed point.\footnote{For higher $q$ this convergence requires much more iterations, more than 36000 for $q=16$. But the values (\ref{pex1})  appearing through the stable phase would be very similar in the $q=14,\,q=16$ simulations.  Indeed as will be explained in section \ref{tnr} these two configurations can be considered to encode the same physical model.}
 Note however that this number of iterations corresponds to an extremely large lattice (in lattice units). 

This type of behavior is typical for fixed points with unstable directions but could also occur for quasi fixed points. To differentiate between these two cases, we considered a one--parameter deformation of the $q=14,K=2$ model, for which we changed the $Q(1)=Q(2)=Q(12)=Q(13)$ values from unity to an arbitrary parameter $0<x<1$. Indeed, there is a phase boundary for $x\sim 0.365$ which leads to a non--trivial fixed point (we give only the first two non-vanishing digits)
\ba\label{pex2}
\textstyle  Q = {\scriptstyle (1,\, 0.72,\,0.28,\,0.057,\,0.0062,\,0.00035,\,0.000011,\,0.00000070,\,0.000011,\,0.00035,\,0.0062,\,0.057,\,0.28,\,0.72) }.
\ea
This fixed point is considerably different from the configuration (\ref{pex1}) and also the initial configurations parameterized through $x=1$ and $x=0.365$ are quite different. Hence we see that a rather large portion of parameter space is dominated by the unstable fixed point. To obtain convergence to either LTF and HTF extremely large iteration numbers are necessary. In other words, the phase transition between LTF and HTF is very weak and the phase boundary not very pronounced. 

Indeed, in two--dimensional systems and in the limit of a  continuous symmetry  group, such as
$U(1)$,  one cannot expect the usual type of second order phase transition between the symmetry
breaking phase (which here would be LTF) and the disordered phase (HTF). This is explained by the
Mermin--Wagner (Coleman) theorem stating that continous symmetries cannot be spontaneously broken at
finite temperature \cite{hohenberg,mermin, coleman}. Nevertheless, there are two phases, both in the
two--dimensional system with global $U(1)$ symmetry \cite{spencer1,spencer2} connected by a
Kosterlitz-Thouless transition \cite{KT} and in the four--dimensional gauge system
\cite{polya,banks}. This is a phase transition of infinite order.  However, for the (isotropic) MK
relations it was proven by Ito (again under certain assumptions on the initial configurations,
including positivity of weights both in group and Fourier space), that this phase transition is not
detected \cite{Ito:1985bv}. But also here extremely slow convergence (of all configurations towards
HTF) has to be expected due to the existence of quasi fixed points. Our findings are explained by
these considerations (although we use initial configurations not satisfying Ito's assumptions), that
is by going to larger $q$ we have to expect weaker and weaker phase transitions. Furthermore, the
Ising type fixed point on the invariant line between LTF and HTF is drawn continuously towards HTF,
indicating the absence of the transition in the limit of large $q$. For 4D lattice gauge theories
with non--Abelian Lie groups it is speculated \cite{Ito:1985bv} that the MK relations provide a
better approximations than for Abelian ones. The confinement conjecture states that in contrast to
Abelian groups there is no phase transition for systems with non--Abelian Lie groups \cite{banks}.

\section{The tensor network renormalization scheme} \label{tnr}

Here we will discuss a second coarse graining method  and apply it to spin net models in 2D in the spin net representation.

We have shown in section \ref{spinnets} that the spin net models can be easily brought into tensor
network form. For these a number of real space renormalization techniques have been developed in 
recent years \cite{levin,Gu:2009dr,levintalk}. Moreover, tensor networks have became
popular not only as a tool to formulate partition functions for `classical statitistical models' but
also to provide a variational ansatz for trial wave functions \cite{Cirac,vidalprl,gulevinwen} for
quantum statistical models. For a variational ansatz one has to find the expectation value of the
Hamitonian with respect to the trial wave functions. The computational techniques \cite{gulevinwen} 
are similar to the tensor network renormalization group techniques. 

This is another point that motivates us to consider renormalization of spin net models, as structures
very similar to spin nets, the so-called spin networks, appear in the canonical or Hilbert space
formulation of spin foam models. Hence the renormalization techniques considered here could also be
useful to find e.g.\ the physical wave function (ground state) via a variational ansatz. Indeed, the
tensor network ansatz has also been developed to describe topological phases
\cite{Gu:2009dr,gulevinwen}, which often are represented by the so-called physical wave functions of
$BF$--theories, which are the starting point of spin foam quantization.  Moreover, as we have also
seen in section \ref{spinnets}, tensor networks and spin networks  are naturally related
\cite{penrose2,Singh:2009cd}. More generally, tensor networks are a general tool for graphical calculus \cite{penrose2,biamonte},
which becomes especially powerful for representation theoretic models such as spin foams and spin
nets \cite{penrose1,Barrett:2009mw}.

The starting point of the tensor network renormalization (TNR) method is to write the partition function of a given model as a contraction of tensors associated to vertices of a graph (or lattice)
\ba\label{au1}
Z=\sum_{a,b,c,d\ldots} T^{abcd} T^{aefg} T^{bhij} \cdots  \q .
\ea
The contraction of indices is along the edges of the graph, that is every edge of the graph carries  one index. As these are summed over we can interpret the indices as the variables or degrees of freedom of the model. The dynamics is encoded in the choice of tensor. 

The tensor network itself, that is its underlying graph, can also be interpreted as a lattice in
space or space--time.  Choosing appropriate subsets of tensors and contracting all tensors inside
each subset according to the connectivity given by the network will result in a network with fewer
vertices and edges. Hence, this procedure corresponds to the blocking procedure in real space
renormalization. Each subset results in a new `effective' tensor $T'$ describing an effective model.
Notice, however, that in the two and higher dimensional case  the effective tensors $T'$
generically\footnote{An exception are hierarchical lattices \cite{Ito:1984ph}.} carry more indices
than the original tensors $T$. For a regular lattice the number of indices  associated to the
effective tensors $T'$ grows with the number of iteration steps. For the underlying graph it means
that the valency of the effective vertices will grow --  mostly in the form of having multiple edges
between pairs of vertices. These multiple edges can be summarized into effective edges, which then
carry indices with an exponentially growing range.

Here is where one has to choose an approximation such that the indices run over a pre-chosen maximum
number of values $D_{c}$. By increasing $D_{c}$ the approximation can be improved systematically.
Ideally, this approximation should pick out only the relevant physics and neglect the irrelevant
short distance fluctuations. The details of how this selection is implemented depend on the scheme,
here we will follow a refined version of \cite{levin,Gu:2009dr}.

The TNR method is very general as many different models can be written in tensor network form.
That is in principle one can also flow between different models  based on different kind of
variables.  
models we obtain the same initial tensor network models for which also follow the same
renormalization flow. 
On the other hand one loses a direct physical interpretation of the 'blocking procedure', i.e.\ how
the effective/ blocked degrees of freedom are built up from the microscopic ones. This information
can be supplemented by studying expectation values of (coarse grained) observables. Below we will
introduce a method which allows to keep some physical interpretation of the indices associated to
the effective tensors. This is however specific to the spin net/ spin foam representation, where
the indices are group representation labels and for the gravitational spin foam models carry
geometric information, such as lengths and area values. Hence we can argue that this method would
correspond to blocking over these area or lengths variables. Such a blocking procedure was also
employed in \cite{Bahr:2009qc, Bahr:2010cq} for the classical Regge model and has the advantage of a
direct geometrical interpretation of the coarse graining procedure.

 Here we will consider the TNR method for a regular lattice but the principle is also applicable to generic graphs arising for instance from random triangulations or Feynamn graphs. In this case one needs, however, to find some suitable approximation scheme to prevent an exponentially growing index range of the effective tensors.

\subsection{Gau\ss~constraint preserving TNR method }

In this section we will shortly describe the TNR method following \cite{levin,Gu:2009dr} applied to
2D Abelian spin net models. We will, however, introduce a technique to keep the Gau\ss~constraints
explicitly valid throughout the renormalization process, see also \cite{Singh:2009cd,singh2}. The
reason for doing this is that the Gau\ss~constraints have an immediate geometrical information: In a
given spin net (with oriented edges) consider any region such that its boundary cuts only through
edges. Then only those configurations will contribute to the   partition sum for which the sum of
all ingoing indices is equal (modulo $q$) to the sum of all outgoing indices. This means that the
Gau\ss~constraints should also hold at the effective vertices, which arise from blocking all the
vertices in certain regions.
We will first review the method for a general 2D tensor network model based on a square lattice and
afterwards specify to the case of spin net models and deal with the Gau\ss~constraints.

\begin{figure}
\begin{center}
 \mbox{
      \subfigure[square lattice]{\includegraphics[width=0.45\textwidth]{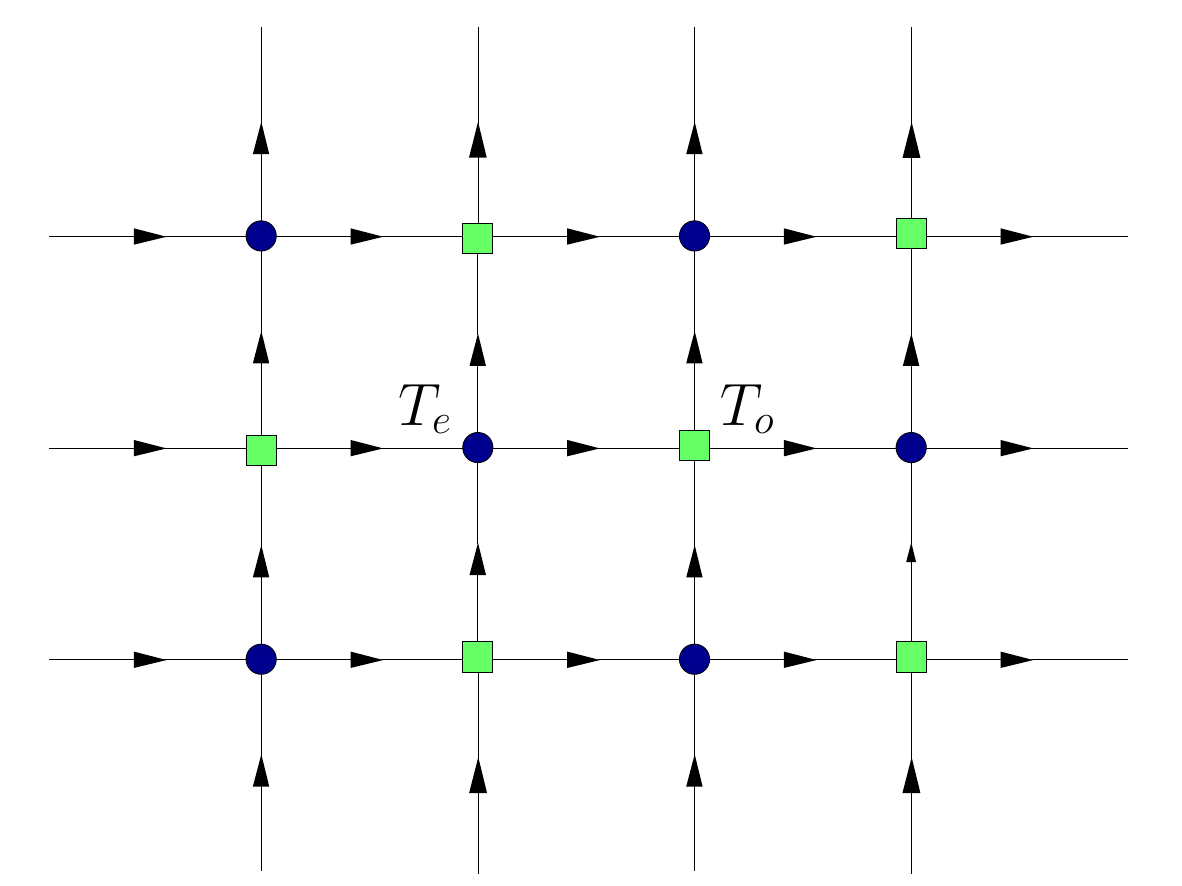}} \qquad
      \subfigure[splitting of vertices]{\includegraphics[width=0.45\textwidth]{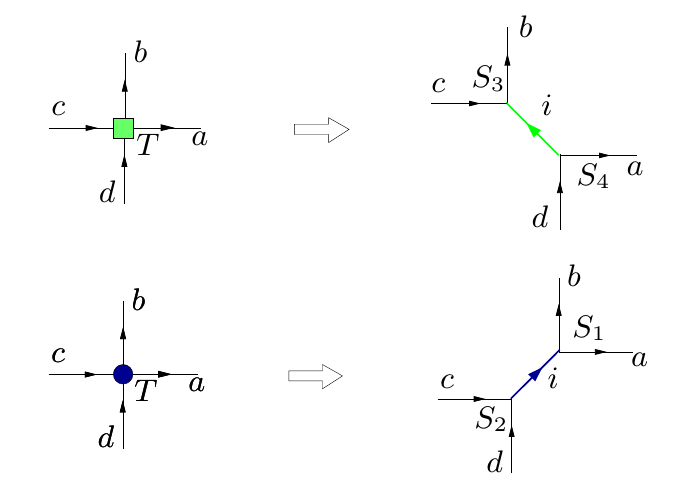}}
      }
%
\vspace*{0.2cm}
\caption{The square lattice with even (blue circles) and odd (green squares)  vertices.
\label{lattice}
}
\end{center}
\end{figure}

Consider a 2D tensor network based on a square lattice, so that the tensors $T^{abcd}$ are of rank
four, see figure \ref{lattice}a. An obvious way to proceed would be to contract always four tensors
along a square and to define in this way a new effective tensor which would now carry four double
indices. 

However, to find a suitable approximation, that is a method to keep the index range constant, one proceeds differently. The first step 
 is to decompose the tensors $T$ into a product of two other tensors $S$. This is performed in two
different ways according to the partition of vertices into odd and even ones. A vertex is even,
respectively odd, if the sum of its lattice coordinates is even, respectively odd. 
 
%

For even vertices we decompose  (see figure \ref{lattice}b)
\ba\label{t2}
T^{abcd}=\sum_i S_1^{ab,i} S_2^{cd,i}  \q .
\ea
Such a decomposition is always possible using a singular value decomposition (SVD) for the
$d^2\times d^2$ matrix $M_1^{ab,cd}=T^{abcd}$. Here $d$ gives the range of the indices $a,b,\ldots$.
This gives 
\ba\label{t3}
M_1^{ab,cd} = \sum_{i=0}^{q^2-1} U_1^{ab,i} \lambda_i (V_1^\dagger)^{i,cd}
\ea
with positive singular values $\lambda_i$ and unitary matrices $U$ and $V$. We  can then define $S_1^{ab,i}=\sqrt{\lambda_i} U_1^{ab,i}$ and $S_2^{cd,i}= \sqrt{\lambda_i} (V_1^\dagger)^{i,cd}$.

Similarly for the odd vertices we decompose (see figure \ref{lattice}b)
\ba
T^{abcd}= \sum_i S_3^{cb,i} S_4^{ad,i}
\ea
where now one uses a SVD for the matrix $M_2^{cb,ad}=T^{abcd}$.

\begin{figure}
\begin{center}
 \mbox{
      \subfigure[contraction]{\includegraphics[width=0.60\textwidth]{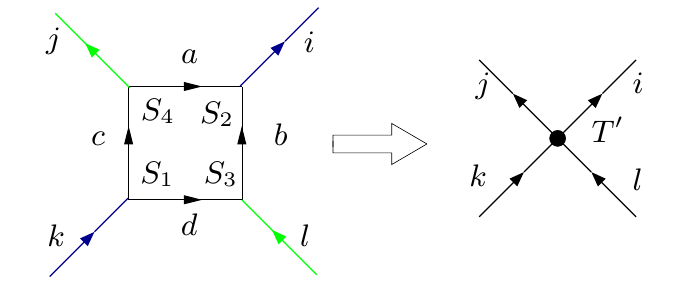}} \qquad
      \subfigure[coarse grained lattice]{\includegraphics[width=0.24\textwidth]{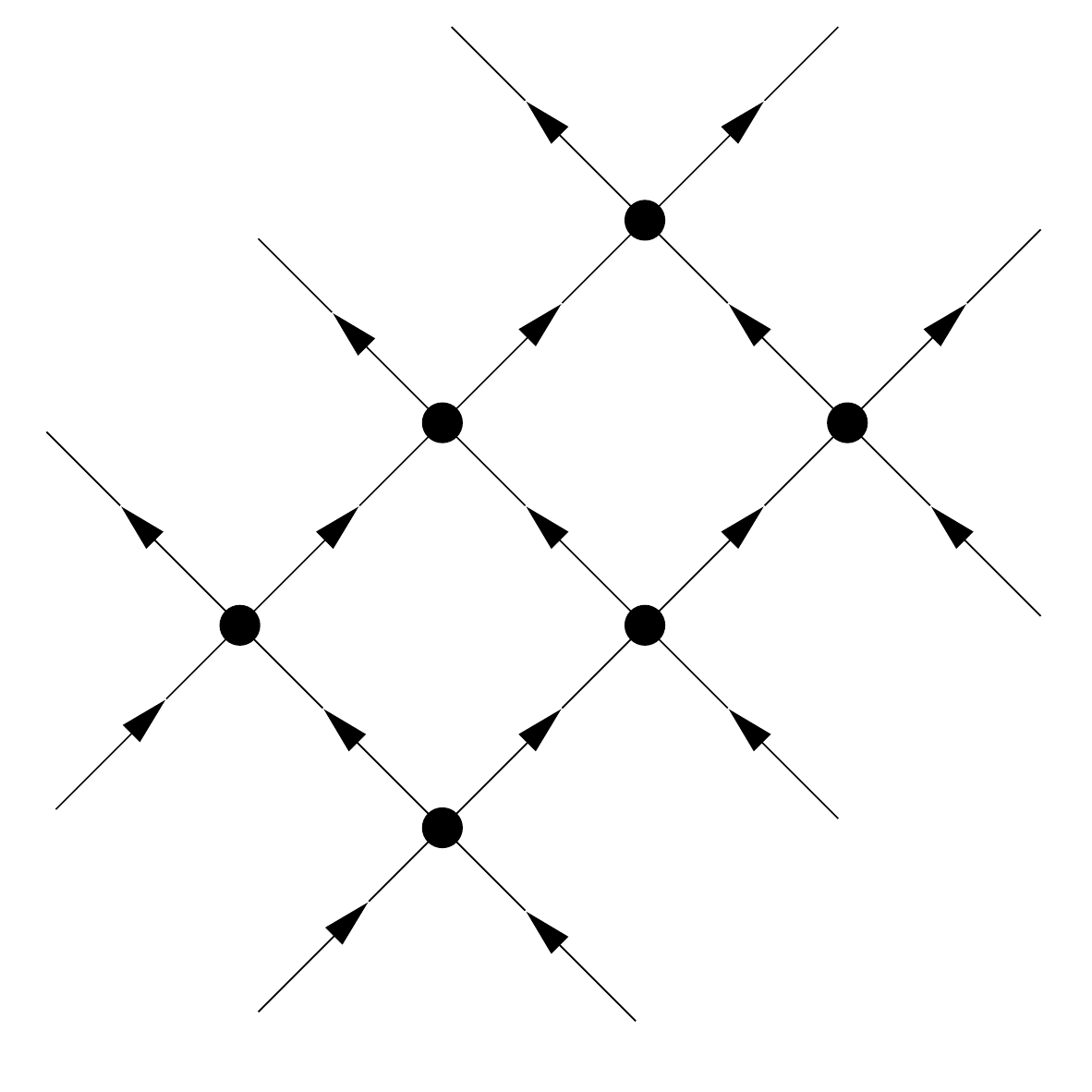}}
      }

\caption{$(a)$ Contraction of the four $S$ tensor to the new $T'$ tensor. $(b)$ The coarse grained lattice.
\label{fig:cvertex}
}
\end{center}
\end{figure}

 In a second step we contract four of the tensors $S$ along the indices of type $a,b,\ldots$ to
obtain the new tensor $T'^{ijkl}$, now with indices $i,j,\ldots$ (see figure \ref{fig:cvertex}a), and
arranged along a square lattice rotated by $45^\circ$ (see figure \ref{fig:cvertex}b)
 \ba
 {T'}^{ijkl}=\sum_{a,b,c,d} S_2^{ab,i}S_4^{ac,j} S_1^{dc,k}S_3^{db,l} \q .
\ea

 If we keep the range of $i$ as in equation (\ref{t3}) the index range of the tensors $T$ would grow exponentially with the number of iterations. This is where the key approximation step comes in, namely to consider only the $D_c$ largest singular values in the decomposition (\ref{t3}). This approximation is justified as the partition function is a trace over the tensors, thus involving the sum over the singular values. The validity of the approximation can be checked by comparing the values of the neglected singular values against the largest singular values in the SVD \cite{levin}. One can choose a rescaling after each iteration step such that this largest singular value is equal to one. Implementing the cutoff $D_c$ in the number of singular values in the decomposition (\ref{t3}) we will obtain a flow in the space of tensors of rank four with a constant index range given by $D_c$. 
 
 The SVD does not only serve as an approximation method but leads also to a field redefinition. Here
the field variables are given by the indices over which the tensors are contracted. In the SVD these
tensors are linearly transformed, which also induces a transformation on the fields. The
transformations aim at an efficient representation of the partition sum, i.e.\ involving a minimal
range of indices or equivalently minimal number or range of variables. The SVD is not unique, in
particular for degenerate singular values one can add rotation matrices acting on the eigenspaces
associated to the degenerate singular values.  For the spin net models we will, however, modify the
method so that the kind of field redefinitions that can occur are restricted. This is related to
preserving the Gau\ss~constraints throughout the renormalization process. It also has the advantage
that the indices keep their original physical interpretation, which for the gravitational models are
related to geometrical quantities.

 Let us now specify to the Abelian spin net models in the spin net representation. Here it is
convenient to introduce an orientation for the edges: for the square lattice we will choose all
horizontal edges to point to the right and all vertical edges to point upwards.
In this case the initial tensor $T$ is of the form (the indices are anti-clockwise cyclically
ordered starting from the leg pointing to the right, as in figure \ref{lattice}b) 
 \ba\label{ff1}
 T^{abcd}=u(a)u(b)u(c)u(d)\,\, \delta^{(q)}(a+b-c-d) \q 
 \ea
where $u(\cdot)=\sqrt{\tilde w(\cdot)}$ in the notation of (\ref{Aco}) and $a=0, \ldots, q-1$ for a model based on $\Zl_q$. The delta function factor signifies the Gau\ss~constraints. Because of this factor the matrices $M_1$ and $M_2$ can be brought into block diagonal form, namely for $M_1^{ab,cd}$ we have the condition that 
 \be
 a+b=c+d=:i \q\q \text{mod}\q q
 \ee
 for non-vanishing entries, whereas for $M_2^{cb,ad}$ we have that 
 \be
 b-c=d-a=:j
 \ee
  Here the indices $i,j$ label the non--vanishing blocks. 
 
In the first iteration step the decomposition into the tensors $S$ can be obtained exactly and
involves at most $q$ non--vanishing singular values 
 \begin{align}
 &S_1^{ab,i}=u(a)u(b)\,\delta^{(q)}(a+b-i) \;,\q&& S_2^{cd,i}=u(c)u(d)\,\delta^{(q)}(c+d-i)\; , \nn\\
  &S_3^{cb,j}=u(c)u(b)\,\delta^{(q)}(b-c-j) \; ,\q && S_4^{ad,j}=u(a)u(d)\, \delta^{(q)}(d-a-j) \q .
\end{align}
We assume that $D_c>q$ (or alternatively that $D_c$ is bigger than the number of non--vanishing $u(a)$), so that no approximation is necessary at the first iteration step. Note that at least $D_c=q$ is necessary to flow to the low temperature fixed point, where $u(a)=1$ for $a=0,\ldots,q-1$. (As one can check the corresponding tensor is a fixed point also for the tensor network renormalization flow.)

 The contraction of four tensors $S$ along the four edges of the square would involve four sums. Due
to the (four) delta functions in the sum this, however, reduces to one summation. There will be one
delta function $\delta^{(q)}(i+j-k-l)$ left, which amounts to the Gau\ss~constraint for the
effective tensor $T'^{ijkl}$:
 \ba
 T'^{ijkl}= \delta^{(q)}(i+j-k-l) \, \sum_c \, u^2(c) \, u^2(i-c) \, u^2(j+c)\, u^2(k-j-c) \q .
 \ea
 
 This new tensor will in general not be of the factorizing form (\ref{ff1}) anymore, so generically
the decomposition into tensors $S$ will now involve $q^2$ non--vanishing singular values. That is in
case $D_c<q^2$, the approximation sets in. Nevertheless, the block diagonal form of all tensors and
matrices involved can be kept through the following iterations. 
 
At a general iteration step we will work with a tensor $T^{abcd}$ with double indices $a=(j_a,m_a)$
where $j_a=0,\ldots q-1$.  As will be explained below the range of $m_a$ depends on the SVD in the
previous iteration step. These tensors will satisfy the Gau\ss~constraints 
 \ba\label{aug1}
 T^{(j_a,m_a)(j_b,m_b)(j_c,m_c)(j_d,m_d)}\sim \delta^{(q)}(j_a+j_b-j_c-j_d) \q 
 \ea
 at all iteration steps.

Similarly as before we can define for the even vertices the matrix
$M_1^{(j_a,m_a)(j_b,m_b),(j_c,m_c)(j_d,m_d)}= T^{(j_a,m_a)(j_b,m_b)(j_c,m_c)(j_d,m_d)}$. Due to the
Gau\ss~constraint (\ref{aug1}) this matrix can be brought into block diagonal form, as the
non--vanishing entries must verify $j_a+j_b=j_c+j_d=:i$. We will denote these blocks by $M_1(i)$.
Then the singular value decomposition can be applied on the single blocks $M_1(i)$ for
$i=0,\ldots,q-1$, so that
\ba
(M_1(i))^{(j_a,m_a)(j_b,m_b),(j_c,m_c)(j_d,m_d)}= \sum_{m_i} (U_{1}(i))^{(j_a,m_a)(j_b,m_b),(i,m_i)}
 \lambda_{m_i} (V_1(i)^\dagger)^{(j_c,m_c)(j_d,m_d),(i,m_i)}\, .\q
\ea

 This yields of course the same singular values as for the entire matrix. Apart from providing a faster algorithm \cite{Singh:2009cd,singh2}, the numerical implementation leads also to more stable results for the following reason. Generically one  encounters the case of having singular values with multiplicities higher than one. In this case the singular value decomposition is not unique as one can perform rotations among the basis vectors associated to a given singular value. Here, keeping the block structure explicit prevents a mixing between different blocks induced by these rotations.
 
To implement the approximation we now have to select the $D_c$ largest singular values among the
singular values of each block.  The number $N_1(i)$ of singular values selected from the block $i$
determines the range of the second index $m_i$ in the double index $(i,m_i)$. That is this number
$N_1(i)$ can take values between zero (no singular value selected) and $D_c$ (all selected singular
values come from one and the same block). Note that for the first iteration step $N_1(i)=1$ (in case
all the $u(a)$ are non--vanishing). We define the $S$ matrices by
\ba
S_1^{(j_a,m_a)(j_b,m_b),(i,m_i)}&=& \sqrt{\lambda_{m_i}} (U_{1}(i))^{(j_a,m_a)(j_b,m_b),(i,m_i)}  \,\, ,\nn\\  S_2^{(j_a,m_a)(j_b,m_b),(i,m_i)}&=& \sqrt{\lambda_{m_i}} (V_1(i)^\dagger)^{(j_c,m_c)(j_d,m_d),(i,m_i)}  \q .
\ea
Note that we have $i=j_a+j_b$ and $i=j_c+j_d$ (mod $q$) for the non--vanishing entries of $S_1$ and $S_2$ respectively. 
 For the matrices  $(M_2(i))^{(j_c,m_c)(j_b,m_b),(j_a,m_a)(j_d,m_d)}$ at the odd vertices we proceed similarly, which will result in matrices $S_3^{(j_c,m_c)(j_b,m_b),(i,m_i)}$ and $S_4^{(j_a,m_a)(j_d,m_d),(i,m_i)}$.  For the non--vanishing entries of these matrices we have $j_b-j_c=i$ and $j_d-j_a=i$ (mod $q$) respectively. The range of the indices $m_i$ is now determined by the number $N_2(i)$ of singular values selected from the block~$i$ of the matrix $M_2$.
 
 The contraction of the four $S$--matrices along the square now results  in a sum
 \ba
 {T'}^{(i,m_i)(j,m_j)(k,m_k)(l,m_l)}&=&
\sum_{c}\sum_{m_c,m_{i-c},m_{j+c},m_{k-j-c}} S_2^{(c,m_c)(i-c,m_{i-c}),(i,m_i)} S_4^{(c,m_c) (j+c,m_{j+c}),(j,m_j)} \nn\\
&&\q\q S_1^{(k-j-c, m_{k-j-c}) (j+c,m_{j+c}),(k,m_k) } S_3^{(k-j-c,m_{k-j-c})(i-c,m_{i-c}),(l,m_l)}  \nn\\
&\sim& \delta^{(q)}(i+j-k-l) \q .
 \ea
Here the range of  $m_c, m_{k-j-c}$ is determined by $N_1(c),N_1(k-j-c)$ from the previous iteration
step whereas $m_{i-c},m_{j+c}$ is determined by $N_2(i-c),N_2(j+c)$ respectively, also from the
previous iteration step. Accordingly, the range of $m_i,m_k$ is determined by $N_1(i),N_1(k)$ of the
iteration step under consideration (that is by the last SVD) and $m_j,m_l$ by  $N_2(j),N_2(l)$
respectively.

This altered algorithm has not only the advantage of keeping the Gau\ss~constraint explicit but is
also keeping any physical interpretation that might be attached to the representation labels~/
indices $j_a$. For the gravitational spin foam models these would carry information on lengths and
area variables -- and the procedure here would correspond to a blocking where the microscopic
geometrical variables are basically added to obtain the coarse grained variables. The
Gau\ss~constraints arise because of reasons rooted in representation theory: namely that for Abelian
groups the tensor product of representations $k$ and $k'$ leads to a representation $k+k'$.  Indeed
the tensors $T$ and $S$ can just be seen as intertwining maps between (tensor products of)
representation spaces, and the first index $j_a$ of the double index $(j_a,m_a)$ encodes the
representation carried by the associated edge.

The same arguments based on representation theory apply for the non--Abelian spin net models.
Also there the tensors $T$ and $S$ are intertwining maps between representation spaces leading to
matrices of block diagonal form. For the abstract spin net models the initial tensor $T$ is
basically determined by the choice of projector $\tilde P^v$ in (\ref{tprod}), which restricts the
intertwining map from one of maximal rank to one of some smaller rank. Here it will be interesting
to study whether any of the projector properties are preserved under coarse graining. 

Tensor network methods thus have the potential to give direct insight in the behavior of the
projectors $\tilde P^v$ under coarse graining, which are the key dynamical entities in spin foam
models. Moreover, the kind of coarse graining in the spin net or spin foam representation uses the
representation theory underlying these models and furthermore corresponds to a geometrically natural
coarse graining.

\subsection{Equivalence of models}\label{secequi}

A crucial point in every renormalization method is the question  which space of models one is
considering, that is in which space the renormalization flow is taking place.  Often the coarse
graining process leads to models outside this space and usually some approximation method is
employed to project the models back into the chosen space. For instance, in the Migdal-Kadanoff
approach one considers models with local interactions, that is non-local terms have either to be
neglected or replaced by appropriately chosen local interaction terms. In particular for a $\Zl_q$
gauge model with local (single plaquette) interactions one always stays in the form of the $\Zl_q$
gauge model (with single plaquette interactions).

In contrast the space of models in the TNR method is defined by the chosen cutoff $D_c$ on the index
range of the tensors $T$. Hence different models, for instance spin net models with different
$\Zl_q$ groups, can be considered in the same space. Indeed, for certain initial conditions the
initial tensors $T^{abcd}$ might actually define the same tensor network models.

Consider, for instance, the Abelian cutoff models for different (even) $q$ but for the same cutoff parameter $K$. That is we deal with the initial tensor (\ref{ff1}) where 
\ba\label{tues1}
u(k)=
\begin{cases}
&1,\q\text{for}\; |k|\leq K \\
&0,\q\text{for}\;|k|>K
\end{cases}
\ea
with $k$ running from $-(\tfrac{q}{2}-1)$ to $\tfrac{q}{2}$ (we consider even $q$).

As $T^{abcd}=u(a)u(b)u(c)u(d)\delta^{(q)}(a+b-c-d)$ the corresponding (symmetric) matrices $M^{ab,cd}_1$ and $M^{bc,ad}_2$ will have  zero rows and columns if these include an index $a$ with $u(a)=0$. In the SVD these rows/columns can just be neglected as it leads to a vanishing singular value. After neglecting zero rows and columns in the matrices arising from different models, these matrices might however coincide (with appropriate matching of indices) and in this sense define the same TNW model. 

Indeed, one can check that in the case of the models (\ref{tues1}) this occurs for a fixed cutoff $K$ but for varying $q$ as long as $q \geq 4K+2$. In this case the first singular value decomposition for the matrices $M_1,M_2$ give for both matrices the same $4K+1$ non--vanishing singular values 
\be
\lambda_{max},\lambda_{max}-1,\lambda_{max}-1,\lambda_{max}-2,\lambda_{max}-2,\ldots,1,1 \q , \q\text{where}\; \lambda_{max}=2K+1\q .
\ee
As long as the cutoff in singular values $D_c$ is chosen such that $D_c\geq4K+1$ the first decomposition (\ref{t2}) of the $T$ matrices into $S$ matrices is exact.

In particular this means that  Abelian spin net models  with $q\geq 4K+2$ and fixed $K$ but
different $q$ should go through the same renormalization sequence and hence should also end in the
same fixed point. We will see in section \ref{tnwaco} that this holds almost always in the numerical
simulations. There will however be also examples where the fixed points depend on $q$. In these
cases the simulations approach an unstable fixed point and the difference in the simulations
appear only for large iteration numbers (more than 100 iterations).  
Hence, the difference can be explained by numerical instabilities and the fact that our method
depends on the parameter $q$ in order to keep the block structure of the $T$ matrices explicit: $q$
 defines the number of blocks and hence the kind of possible field redefinitions which underly
the method.

Disregarding this point the equivalence between models should also hold if we send $q \rightarrow
\infty$, that is consider $U(1)$. In general, we see that the TNR method might also be applied to
Lie
groups (which would lead to infinite dimensional matrices) as long as we consider initial data such
that the initial matrices reduce to finite dimensional ones due to either the appearance of zero
singular values, or singular values which are sufficiently small. This equivalence between models
appears only approximately in the Migdal--Kadanoff method, as there the number of parameters on
which the renormalization flow acts is fixed by $q$, or more generally the size of the group, on
which the model is based.

\subsection{Structure of fixed points}\label{secfp}

In the Migdal-Kadanoff scheme we considered a renormalization flow within a space described by $q$
(or $(q-1)$ after normalization) parameters for spin nets based on the group $\Zl_q$. We are much
more flexible with the TNR method, where the number of parameters is determined by the chosen cutoff
$D_c$ on the number of singular values. Hence there are potentially  many more fixed points. Indeed
the fixed point structure is considerably more complicated, in particular for $D_c>>q$, as we will
exemplify below with the Ising model, $q=2$.  Note that (with the exception of $D_c=2$, which can be
treated analytically) we will  only discuss fixed points which we found as a result of the
renormalization process, i.e.\ by flowing to these fixed points. That is we will miss most of the
unstable fixed points, which are those with repellent directions and would require a fine tuning of
parameters to flow into. 

The main feature of  the TNR method -- with the approximation based on the singular value
decomposition  -- is the appearance of non--isolated fixed points. These are argued \cite{levintalk,
Gu:2009dr} to be due to short scale degrees of freedom which are not averaged out by the
approximation method employed here. In \cite{Gu:2009dr} different forms of additional approximation
steps (termed entanglement filtering) are suggested, that apply once the flow reaches the
non--isolated fixed points. We will not consider these additional steps here and just describe the
type of fixed point encountered in the original TNR method. For future work one should, however,
study this issue in more detail. In particular, one could benefit from a detailed comparison between
approximations of Migdal-Kadanoff type and approximations based on the TNR method and furthermore
from an analysis of relevant and irrelevant directions for the linearized renormalization flow
around this kind of fixed points.

In the following we will describe the results of the TNR method for the Ising model in the spin net representation (associated to the high temperature expansion) for the choice of different cutoffs $D_c$, see also \cite{ Gu:2009dr} for a treatment of the Ising model with the TNR method improved by entanglement filtering.  

The smallest cutoff which would also allow a representation of the low temperature fixed point (LTF) $u(0)=u(1)=1$ is $D_c=2$. 
We start with configurations $u(0)=1,u(1)=x$. For $x > 0.60858$ these flow to the LTF represented by
$u(0)=u(1)=1$. Accordingly, the singular values $\lambda_i$ for this fixed point are
$\lambda_0=1,\lambda=1$ with every block $j=0,1$ contributing one singular value.  (We normalize the
tensors in every step such that the largest singular value is equal to one.) 

For $0\leq x < 0.60857$ the configurations flow towards the high temperature fixed point (HTF)
represented by $u(0)=1, u(1)=0$. This corresponds to having only one non--vanishing singular value
$\lambda_0=1$. The transition point at $u(1)=0.6085 $ corresponds to a phase transition temperature
of $kT_c\approx 2.572$. This is a better approximation to the exact result of $kT_c\approx 2.269$  
than the isotropic Migdal-Kadanoff result of $kT_c=3.282$ in section \ref{mkex}. 
 
 In this case, $D_c=2$, one can analytically compute the flow in some suitable parameters, for
instance the tensor components $T^{abcd}$. This allows to find a further (unstable) fixed point,
given by the matrix $M_1^{ab,cd}=T^{abcd}$ (the numbering of the rows and columns is
$ab=00,11,01,10$)
 \ba
 M_1\!=\! \left(
\begin{matrix}
 1+s_2^4 & s_1^2(1+s_2^2) & 0 & 0 \\
 s_1^2(1+s_2^2)  & 2 s_1^4&0&0\\
 0&0& 2 s_1^2 s_2 & s_1^2(1+s_2^2)\\
 0&0&  s_1^2(1+s_2^2) &2 s_1^2 s_2
 \end{matrix}
 \right)  \, ,   \text{with}  \; s_1=0.592902,\; s_2=0.43105.\q
 \ea
Note that the tensor describing this fixed point is no longer of the original form (\ref{ff1})
as this would require $s_2=s_1^2$. Nevertheless it leads to the phase transition between the
low and high temperature regime.

Let us now turn to the case $D_c=4$. In this case configurations with $x>0.6466$ will still flow to the LTF characterized by two non--vanishing singular values $\lambda_0=\lambda_1=1$, one from every block.  For configurations with $x<0.6465$  we encounter however a new type of non--isolated fixed points, so--called Corner Double Line tensors \cite{levintalk,  Gu:2009dr}.

\begin{figure}
 
\begin{center}
\includegraphics[width=0.8\textwidth]{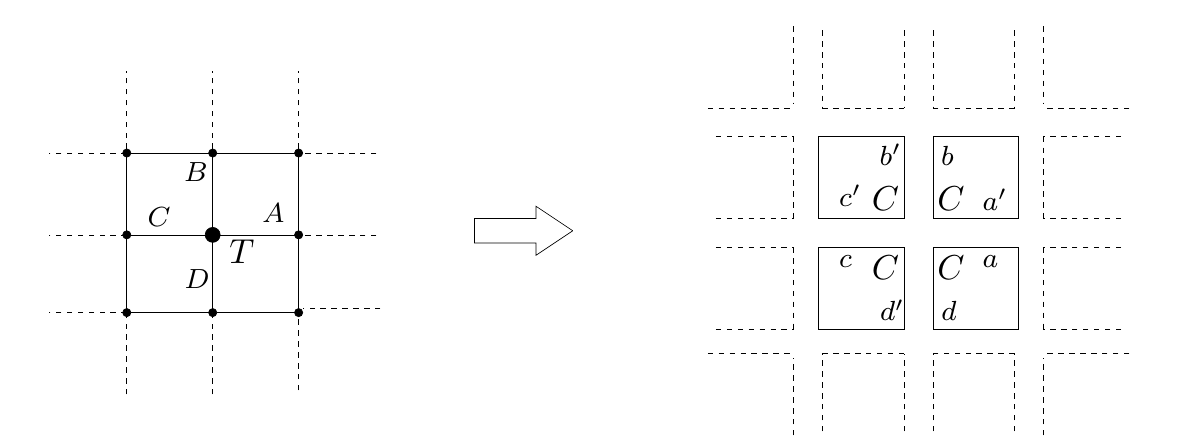}
\vspace*{0.2cm}
\caption{Tensor with Corner Double Line structure
\label{fig:cdl}
}
\end{center}
\end{figure}

`Double Line' indicates that we have to deal with pairs $A=(a,a')$ of indices.  The tensors are
defined by associating a matrix $C^{ab}$ to each corner of the four--valent vertex, see figure
\ref{fig:cdl}, 
\ba\label{don1}
T^{(aa')(bb')(cc')(dd')}=C^{a'b}C^{b'c'}C^{cd'}C^{da}  \q .
\ea
Such tensors are fixed points of the renormalization flow (for $D_c$ sufficiently large). The decomposition (\ref{t2}) of such a tensor is determined by the singular value decomposition of $C^{ab}$.  Namely if 
\ba
C^{ab}=\sum_I  u^{a,I} \,\eta_I\, v^{b,I}
\ea
we can write
\ba\label{don3}
T^{(aa')(bb')(cc')(dd')}= C^{a'b} \left(  \sum_{I,I'} u^{b',I}  u^{d,I'}\eta_I \eta_{I'} v^{c',I}v^{a,I'}\right) C^{cd'}  \q .
\ea
Hence if the SVD of $C^{ab}$ includes $n$ non--vanishing singular values $\eta_I$, we will obtain
$n^2$ non--vanishing singular values of the form $\eta_I \eta_{I'}$ for the matrices obtained from
$T$. 

The fixed point we encounter in the high temperature region for $D_c=4$ is of $CDL$ type with corner matrix
\ba\label{don3a}
C =\left(
\begin{matrix}
 1 & 0 \\
 0 & y
 \end{matrix}
 \right)  \q .
\ea
That is in the SVD for the matrices associated to the tensor $T$ we encounter four non--vanishing singular values $\lambda_i=1,y,y,y^2$.  Here $y$ ranges from zero and seems to get arbitrarily close to $1$ for $u(1)=x$ reaching the transition point at $0.6496562...$. Note that here we encounter a continuous fixed point family of `high temperature type'  ranging from having only one non--vanishing singular value equal to one to having four non--vanishing singular values, with three of these almost equal to one. If we take the transition between this fixed point family and the LTF (with two non--vanishing singular values equal to one) as the phase transition we obtain a critial temperature of $kT_c=2.221$. This again is a better approximation to the exact phase transition temperature than the $D_c=2$ result.

The appearance of these non--isolated fixed points is interpreted \cite{levintalk, Gu:2009dr} to be due to short range entanglement,
which is not filtered out by the renormalization flow. Hence certain microscopic details of the models are remembered and the usual universality property of phase transitions does not apply. One can nevertheless argue that the correspondence between phases and fixed points does hold: here models in a certain phase flow to a certain type of fixed points. Indeed we will see that for $D_c=9$ two different types of non--isolated fixed points appear, representing the low and high temperature regime respectively.
In \cite{Gu:2009dr} additional approximation steps are suggested, designed to filter out this short range entanglement. We will leave the investigation of these and other additional approximation steps for future work.

The appearance of the CDL type fixed points suggest that cutoffs $D_c=n^2$ with $n$ a natural number, might lead to a fast convergence, as these accommodate the $n^2$ non--vanishing singular values of a CDL type fixed points. We therefore will also discuss the case $D_c=9$. Here, both the low temperature phase and the high temperature phase are described by non--isolated fixed points. For the initial value $u(1)=x\leq 0.64293$ we flow to fixed points, described by nine non--vanishing singular values of the form 
\be
\lambda_i =1,\,y,\,y,\,y^2,\,z,\,z,\, yz,\, yz,\, z^2 \q .
\ee
(For $x=0.64293$ we have $y=0.7588$ and $z=0.4200$.) This suggest a fixed point of CDL form based on a $3\times 3$ matrix $C$ with singular values $\eta_I=1,y,z$. This  family of fixed points is continuously connected to the proper high temperature fixed point, where $y=z=0$. 

For $u(1)=x\geq 0.64295$, that is in the low temperature phase, we also flow to a non--trivial fixed point, this time characterized by eight non--vanishing singular values of type
\be
\lambda_i=1,1, y,y,y,y,y^2,y^2 \q .
\ee
(For $x= 0.64295$ we obtain $y=0.6434$. For $x \rightarrow 1$ the parameter $y$ approaches $0$ and we obain the proper low temperature fixed point.) Hence this tensor cannot be of CDL type. Indeed it turns out that it is of the form
\ba
T=T_{CDL}\otimes T_{LTF}
\ea
where $T_{CDL}$ is a tensor based on the corner matrix (\ref{don3a}) and $T_{LTF}$ is the tensor associated to the low temperature fixed point, that is of the form (\ref{ff1}) with $u(0)=u(1)=1$. Therefore eight non--vanishing singular values appear, as products of the four singular values $\lambda=1,y,y,y^2$ for the $CDL$ tensor and the two singular values $\lambda'=1,1$ for the LTF tensor. The appearance of this type of fixed point tensors of product form was conjectured in \cite{Gu:2009dr}. We will encounter more fixed points of this form in the next subsection.

The transition between the high temperature family of fixed points and low temperature family of fixed points corresponds to a critical temperature of $kT_c\approx2.274$, which again approximates the exact result $kT_c\approx 2.269$ better than the $D_c=4$ result ($kT_c\approx 2.221$).


\subsection{Analysis of Abelian cutoff models} \label{tnwaco}

We will now discuss the renormalization behavior of Abelian cutoff models to compare it to the
results obtained with the Migdal-Kadanoff method in section \ref{mkaco}. 
We will consider the same kind of configurations as in section \ref{mkaco}, parametrized by the size
of the group $q$ and by the cutoff\footnote{Let us recall that $K$ refers to the cutoff
introduced in the space of group representations, that actually defines the model, while $D_c$ is
the maximum number of singular values considered when approximating the SVD.} $K$. 

Now, as was pointed out in section \ref{secequi} configurations with the same cutoff parameter $K$ but different $q$ might encode the same tensor network model. To utilize this we will choose the same cutoff $D_c$ ($D_c=16,25$ and equal to $32$ for some models) for different $q$, so that the renormalization flow should also be the same in these cases. This is different from the MK method, where the number of parameters always depends on $q$. 

We have seen in section \ref{mkaco} that the MK renormalization flow for the 2D spin net models  was much more involved than for the 3D spin foam models: For sufficiently large groups $\Zl_q$ configurations would undergo a behavior determined by unstable fixed points and rather weak phase transitions leading to very long convergence times. Therefore we have to expect similar properties to appear in the tensor network renormalization. Furthermore, as is usually the case if configurations approach a region around  phase transitions, the approximation implemented by the cuttoff $D_c$  might get less and less accurate. Indeed, the values of the neglected singular values might become comparably large (of order $10^{-1}$ of the largest singular value, which will always be normalized to one), even for the quite large cutoff $D_c=32$. The TNR method, however, offers  the possibility to increase the cutoff and thereby study the influence of the cutoff on the results obtained.

Another issue that appears during a number of simulations are configurations for which the symmetry
$k \rightarrow -k$ (equivalent to reversing all the edges) is broken. The reason for this  is that
singular values usually appear with a two--fold degeneracy, namely one from a $k$--block, the other
from a $(-k)$--block. The singular values from the block with $k=0$ and $k=q/2$ are an exception to
this rule. Also, quite often singular values appear with even higher degeneracy. Now, taking only a
fixed number of singular values into account, one will quite often dismiss one singular value of a
degeneracy pair and in this way break the edge reversing symmetry. As the number of $k=0$ singular
values, which are taken into account by the cutoff is not known a priori, and might even change
during the iterations, it might be quite difficult to find a cutoff where this issue does not
appear.

In regions `near' phase transitions, that is where convergence takes very long, these non-symmetric
modes typically lead to unstable, oscillating behavior. This also signifies that the influence of
the cutoff is not negligible. In the Migdal-Kadanoff method non-symmetric configurations may only
appear due to numerical inaccuracies and it is quite straightforward to implement a symmetrization
after each iteration step. A similar procedure for the TNR method (which is however not as
straightforward to implement as for the MK method) would probably be very helpful to obtain more
reliable results as well. Here we will interpret an oscillating behavior over large iteration times
as indicating the presence of an unstable or quasi fixed point. 

\begin{figure}
\begin{center}
 \mbox{
      \subfigure[$D_c=16$]{\includegraphics[width=0.45\textwidth]{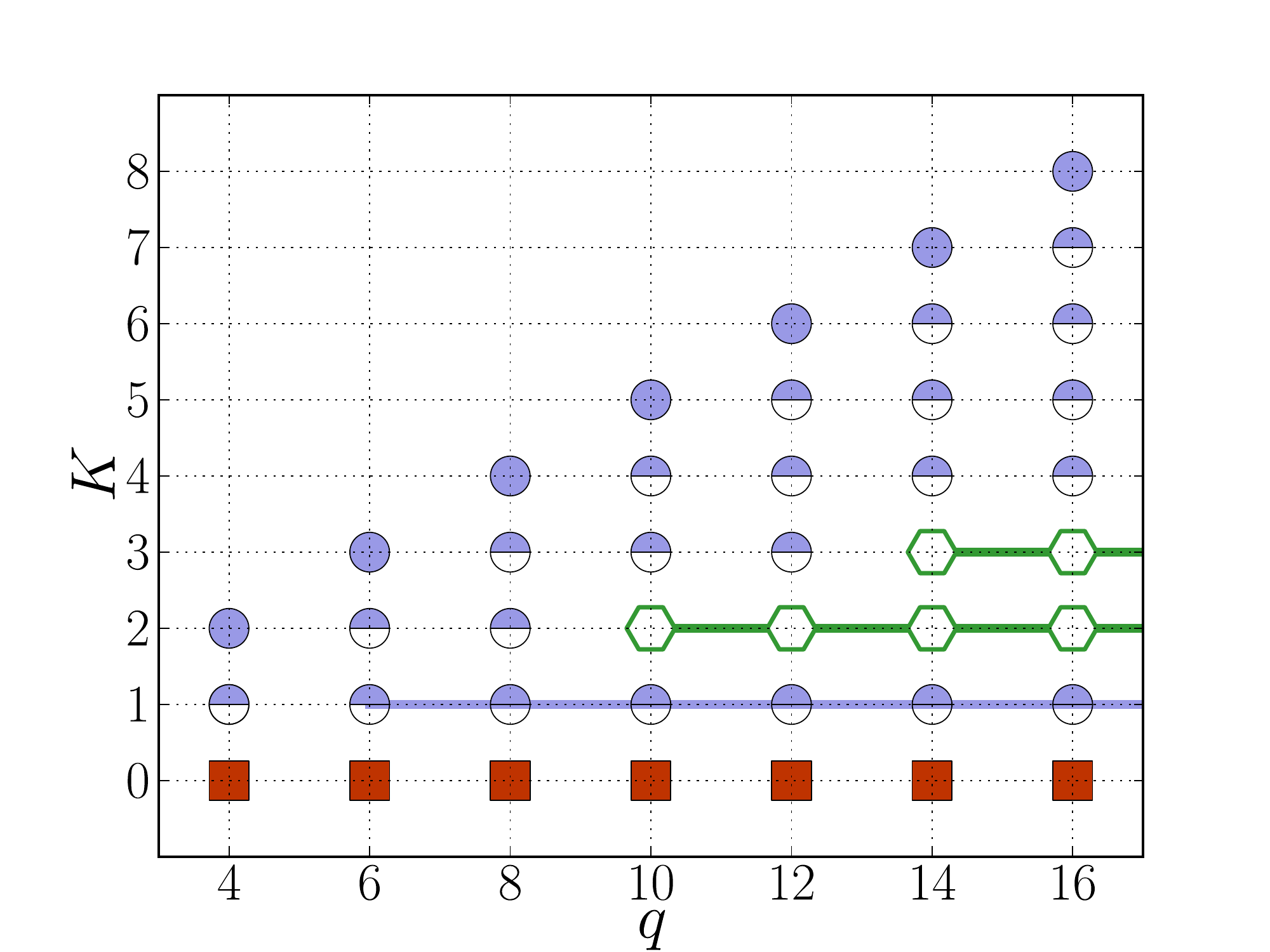}} \quad
      \subfigure[$D_c=25$]{\includegraphics[width=0.45\textwidth]{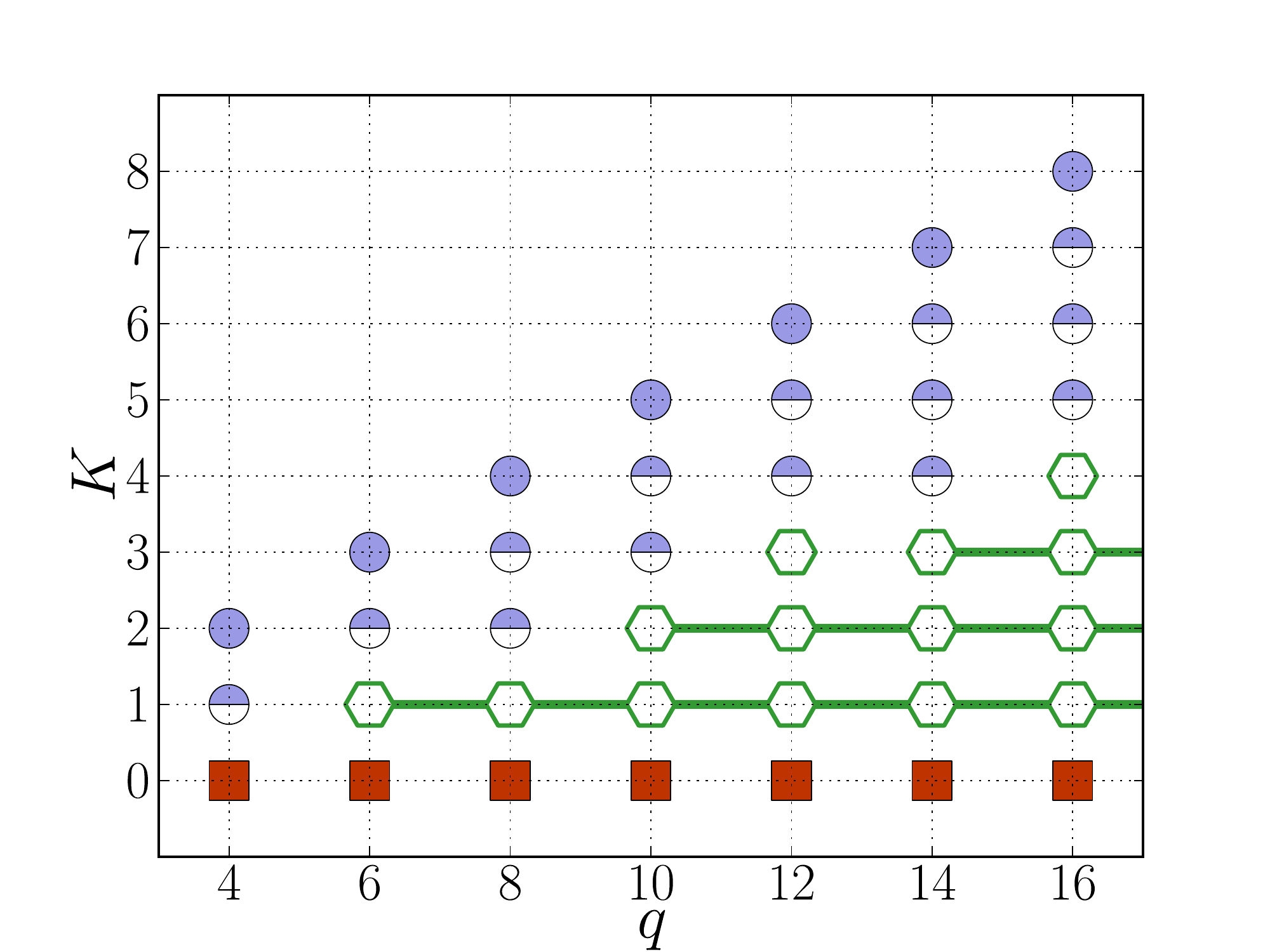}}
     }
\caption{Flow behaviour of different cutoff models, labelled by $K$ and $q$, compare figure \ref{fig:mk-Kq}. Markers half-filled at the top flow to LTF/LTF$\times$CDL (diagonal models, $K = q/2$), empty hexagonal markers indicate a quasi fixed point or oscillating behaviour. Connected markers illustrate equivalent models in the sense of section \ref{secequi}. \label{tnrkq}}
\end{center}
\end{figure}

We can broadly summarize the behavior we encountered in the simulations of the Abelian cutoff models into the following classes, wich we also illustrate in figure \ref{tnrkq}:
\begin{itemize}
\item 
The models flow quite fast (typically during 10 to 30 iterations) to the low temperature fixed point or to a fixed point of type $T_{LTF}\otimes T_{CDL}$ as described in section \ref{secfp}. This happens for cutoff models which are not independent of $q$. The low temperature fixed points come with $q$ non--vanishing singular values equal to unity. 
The more complicated fixed points with CDL structure appear, in particular, for the higher cutoff $D_c=25$. The singular values associated to the CDL structures are however quite small ($\sim 10^{-5}$ to $0.2$), so that we can safely interpret these fixed points as of low temperature type. It also happens that initially the configurations flow to a $T_{LTF}\otimes T_{CDL}$ fixed point with a higher number of non--vanishing singular values than provided by the chosen cutoff. In this case the configurations slowly keep changing such that the singular values associated to the CDL structure decrease. With the exception of two models ($K=3,q=12$ and $K=4,q=16$) such a behavior appearing in the $D_c = 16$ simulation would be confirmed in the $D_c=25$ treatment.
\item
An interesting special case is $K=1$ and $q\geq 6$ (which describes the same tensor network model independent of $q$ as long as $q\geq6$). For $D_c=16$ the fixed point (reached after about 60  iterations) is of the form $T_{LTF}\otimes T_{CDL}$, but the $T_{LTF}$ factor comes with only four singular values equal to unity. The CDL structure leads to 16 non--vanishing singular values, where apart from the four singular values equal to one, the next eight  ones are equal to $0.86$ and the final four equal to $0.73$. 

The $D_c=25$ results are slightly different: due to the appearance of non-symmetric configurations 
as described above, the configurations are oscillating for more than 80 iterations. Then the results
do actually start to depend on $q$: for $q=6$ a $T_{LTF}\otimes T_{CDL}$ fixed point is reached with
six singular values equal to one (so it is really a $q=6$ low temperature fixed point), 12 singular
values equal to $0.81$ and six equal to $0.655$.  Whereas for $q=8$ a fixed point is approached
(around 200 iterations) with eight singular values equal to one, another sixteen are around $0.3$,
plus eight singular values around $0.12$. This gives more than 25 non--vanishing singular values, so
as described above the configuration is slowly changing, decreasing the CDL singular values. Note
however that due to the high iteration numbers involved and the appearance of `non-symmetric'
configurations the results should be taken with some care.

\item
For the  $D_c=16$ simulations there are two examples $K=2$ and $q=10$ or higher, and $K=3$ and
$q=14$ or higher, which show stable behavior and seem to approach a (non--trivial) fixed point for
long iteration times (approx.\ 20 iterations for $K=2$ and approx.\ 100 iterations for $K=3$). The
change in the singular values during this approach can get very small, with instances where the
change in the singular values are of the order $10^{-7}$. There is a large number of non--vanishing
singular values indicating a complex (quasi/unstable) fixed point. After this stable phase
configurations become unstable and `non-symmetric'  showing a slightly oscillating behavior (changes
of the order $10^{-3}-10^{-2}$ from one to the next iteration). For $K=3$ the configurations
converge to $q$--dependent low temperature fixed points (at iterations $230$ for $q=14$ and $180$
for $q=16$), for $K=2$ the behavior remains unstable for very long iteration times. This behavior is
confirmed for $K=2$ with the $D_c=25$ and $D_c=32$ simulations (where now the stable phase is up to
100 iterations). 
For $K=3$, the $D_C=25$ and $D_c=32$ simulations lead to non-symmetric configurations which show
oscillating behaviour for long iteration times. However, the complicated structure of the largest
$q$ singular values agrees with the one of the simulations with smaller cutoff.


\end{itemize}

Note that examples flowing to a high temperature fixed point do not appear. The best candidate would be the $K=1$ configurations (describing the same tensor network models starting with $q=6$ and larger $q$), which however flow for $D_c=16$ to a fixed point with four singular values equal to unity, plus more non--vanishing singular values due to the CDL structure.

In the following table we list the details of our findings and compare them with the results obtained with the Migdal-Kadanoff method. 

\begin{table}[htbp]
\begin{center}
\begin{tabular}{rrlll}
\toprule
$K$ & $q$ & \textbf{MK} & \textbf{TNR} $D_c=16$ & \textbf{TNR} $D_c=25$ \\
\midrule
1 & 4 & LT{\tiny(05)} & LT{\tiny(6)}$\times$CDL{\tiny(7)} & LT{\tiny(6)}$\times$CDL{\tiny($>$100)} \\ \arrayrulecolor{rule} \cmidrule{4-5} \arrayrulecolor{black}
1 & 6 & LT{\tiny(18)} & LT\textbf{4}{\tiny(50)}$\times$CDL{\tiny(53)} & Osc$\rightarrow$LT\textbf{6}{\tiny(42)}$\times$CDL{\tiny(48)} \\
1 & 8 & HT{\tiny(59)} & LT\textbf{4}{\tiny(50)}$\times$CDL{\tiny(53)} & Osc$\rightarrow$LT\textbf{8}{\tiny(98)}$\times$CDL{\tiny($>$100)} \\
\midrule
2 & 6 & LT{\tiny(04)} & LT{\tiny(19)}$\times$NULL{\tiny(53)} & LT{\tiny(07)}$\times$CDL{\tiny(26)} \\
2 & 8 & LT{\tiny(07)} & LT{\tiny(11)}$\times$NULL{\tiny(14)} & LT{\tiny(18)}$\times$CDL{\tiny(19)} \\ \arrayrulecolor{rule} \cmidrule{4-5} \arrayrulecolor{black}
2 & 10 & LT{\tiny(14)} & QF{\tiny(6)}$\rightarrow$Osc{\tiny(13)} & QF{\tiny(8)}$\rightarrow$Osc{\tiny(53)} \\
2 & 12 & LT{\tiny(59)} & QF{\tiny(6)}$\rightarrow$Osc{\tiny(13)} & QF{\tiny(8)}$\rightarrow$Osc{\tiny(53)} \\
2 & 14 & QF{\tiny(3)}$\rightarrow$LT{\tiny(969)} & QF{\tiny(6)}$\rightarrow$Osc{\tiny(13)} & QF{\tiny(8)}$\rightarrow$Osc{\tiny(53)} \\
\midrule 
3 & 8 & LT{\tiny(003)} & LT{\tiny(04)}$\times$NULL{\tiny(06)} & LT{\tiny(04)}$\times$CDL{\tiny(05)} \\
3 & 10 & LT{\tiny(005)} & LT{\tiny(06)}$\times$NULL{\tiny(09)} & LT{\tiny(17)}$\times$NULL{\tiny(38)} \\
3 & 12 & LT{\tiny(008)} & LT{\tiny(10)}$\times$NULL{\tiny(13)} & Osc \\ \arrayrulecolor{rule} \cmidrule{4-5} \arrayrulecolor{black}
3 & 14 & LT{\tiny(013)} & QF{\tiny(18)}$\rightarrow$Osc{\tiny(49)}$\rightarrow$LT\textbf{14}{\tiny(115)} & Osc \\
3 & 16 & LT{\tiny(030)} & QF{\tiny(18)}$\rightarrow$Osc{\tiny(45)}$\rightarrow$LT\textbf{16}{\tiny(093)} & Osc \\
3 & 18 & LT{\tiny(140)} & - & Osc  \\
3 & 20 & QF{\tiny(4)}$\rightarrow$LT{\tiny($>$1000)} & - & Osc\\
\midrule
4 & 10 & LT{\tiny(04)} & LT{\tiny(4)}$\times$NULL{\tiny(6)} & LT{\tiny(07)}$\times$NULL{\tiny(28)}  \\
4 & 12 & LT{\tiny(05)} & LT{\tiny(6)}$\times$NULL{\tiny(8)} & LT{\tiny(08)}$\times$NULL{\tiny(15)}  \\
4 & 14 & LT{\tiny(06)} & LT{\tiny(7)}$\times$NULL{\tiny(9)} & LT{\tiny(10)}$\times$NULL{\tiny(12)} \\
4 & 16 & LT{\tiny(09)} & LT{\tiny(9)} & Osc   \\ \arrayrulecolor{rule} \cmidrule{4-5} \arrayrulecolor{black}
4 & 18 & LT{\tiny(13)} & - & Osc \\
 \bottomrule

\end{tabular}

\caption{Summary of the renormalization flow of Abelian cutoff models. To a precision of $10^{-10}$,
the numbers in brackets give the iteration times (halfed for TNR to make comparable) it takes to
reach a certain behaviour: LT (low temperature fixed point, bold numbers denote the number of ones
that appear explicit), HT (high temperature fixed point), QF (quasi fixed point), Osc (oscillating
behaviour), CDL (corner double line), NULL (finite part that dies off). For a given $K$ and growing $q$, the fine line indicates the beginning of physically equivalent models in the sense of section \ref{secequi}.}
\end{center}
\end{table}

The overall picture of the Migdal Kadanoff results is confirmed by the  
tensor network simulations: Most configurations flow to the low  
temperature fixed point or to a low temperature fixed point  
embellished with a CDL structure.
With growing $q$, cutoff models with sufficiently small $K$ show long  
stable phases where a fixed point seems to be approached but then enter
an unstable phase with (slightly) oscillating behaviour.  
For some cases these converge finally to a LTF fixed point, however  
this result has to be taken with some care, due to the oscillating  
phase in which `non-symmetric' configurations appear.

There is one important difference between the MK and the TNR results.  
As we explained in section \ref{secequi} examples with $q\geq 4K+2$  
should encode the same physical models both exactly and in the  
approximation provided by choosing the cutoff $D_c$. This property is  
inherent in the TNR method (and only appears to be violated for large  
iteration numbers due to the oscillating behavior). But in the MK  
method the truncation is determined by $q$ and hence the  
renormalization flow is still different for $q$ not much larger than  
$q_K:=4K+2$. For much larger $q$ the renormalization sequences will  
turn out to be almost the same, however this behavior sets in later  
than in the TNR examples. Moreover from \cite{Ito:1985bv} we have to expect  
that going to $q \rightarrow \infty$ the configurations should flow to  
the high temperature fixed point.

This is the reason why the quasi fixed point behavior has to set in  
earlier for the TNR method, as here the results starting with $q_K$  
should in principle also hold in the limit $q\rightarrow \infty$. From  
this point of view it is interesting that we have not seen an example  
in the TNR method (especially not $K=1$, as opposed to the MK  
$K=1,q=8$ example) which would flow to the high temperature fixed  
point. Hence the TNR method might be able to detect two different  
phases for the $U(1)$ theory.

As we have also seen definite conclusions are very much hindered by  
the appearance of non-symmetric configurations, i.e.\ where the $k$-- 
blocks would differ from the $-k$--blocks.  These also appeared in the  
MK simulations for the examples which go through very long almost  
stable phases. In this case non-symmetric configurations only appeared  
due to numerical errors and this problem can be easily cured in the MK  
method by symmetrizing the $Q(k)$ parameters after each coarse  
graining step.

In the TNR method  non-symmetric configurations not only appear due to  
numerical errors but are also caused by the cutoff, which might   
neglect one of a pair of degenerate singular values. This happens  
quite generically if the cutoff is increased (apart from the $D_c=25$  
simulations we also tried $D_c=32$ and inbetween values) and it is  
increasingly difficult to find a value for $D_c$ where this does not  
appear, say, for the first twenty iterations.

This non-symmetric behavior does not matter so much for configurations  
which would converge fast to some stable fixed point but do cause  
long sequences of oscillating behavior for configurations, which we  
suspect would otherwise rather approach slowly an unstable fixed point.
For these examples the ordered sequence of singular values at a given iteration decreases rather  
slowly. Even for $D_c = 32$ the neglected singular values can be of order $10^{-1}$ and an unsymmetric cutoff will have considerable influence on the overall behaviour. This problem is especially pronounced for the 2D  
models with larger $q$, as we there encounter  rather weak phase  
transitions.

For future work one should address this issue\footnote{It does however  
only appear for groups where the dual representation $\rho^*$ is not  
equivalent to the original representation $\rho$. For the rotation  
group $SU(2)$ we rather have $\rho^*\equiv \rho$, but for $U(1)$ we  
have $\rho_k^*\equiv \rho_{-k}$.}. There are different possibilities,  
one is to use a symmetric parametrization in the coarse graining  
procedure, i.e.\ only work with the $k$--blocks where $k\leq \frac{q} 
{2}$. Another option is to use an adaptive cutoff $D_c$ for each  
iteration step, such that it avoids to cut between $(k,-k)$ pairs.

A third option would be to change the approximation scheme slightly by  
choosing a block dependent cutoff $D_c(k)$. This would actually  
simplify the algorithm considerably. For sufficiently high $D_c$ and  
$D_c(k)$ these schemes should yield equivalent results, this has  
however to be tested.

Note that non-symmetric configurations can however also appear due to  
numerical instabilities, so that one might have to implement some  
symmetrization procedure (if one does not work with a parametrization  
that only allows symmetric configurations).

\section{Discussion and outlook}\label{out}

Extracting large scale physics from spin foam models is one of the  
most pressing issues for the field. 
Here we advocated the development and use of coarse  
graining and renormalization techniques so that hopefully a feasible method for 4D models can be found. 
To this end, we introduced a wide range of simplified models,  
spin foams with finite groups and moreover spin nets, which can be  
seen as a dimensionally reduced version of spin foams. 
This not only  
enables us to test and develop coarse graining methods but also to obtain  
some physical insights and to put forward conjectures about the  
dynamical behavior of the full models.

This strategy allows us to adopt coarse graining methods from lattice  gauge theory and condensed matter system, here the Migdal-Kadanoff scheme and the tensor network renormalization method.  
The two methods have different drawbacks and advantages.
The Migdal-Kadanoff scheme facilitates quick results (even a large number of iterations can take only seconds on a PC), so that an overview of the phase structure  encoded in the model can easily be obtained. 
Here the main question is how this method can be generalized to non--Abelian spin foams with non--trivial projectors, which no longer fall into the class of standard lattice gauge theories.

The tensor network renormalization method has the advantage of  
providing a systematic improvement on the accuracy of the results.
The required effort is considerably larger (100 $D_c=32$ iterations may take several days on a PC). The method is however very general and in its version based on the spin net representation, allows direct access  
to the behavior of the (vertex) projector under coarse graining.  
Moreover, the blocking of variables is very natural if one takes into  
account the geometric meaning of the representation labels in the  
gravitational spin foam models. We presented an algorithm  in  
which the Gau\ss~constraint are kept explicitly intact.

The tensor network renormalization method is easily generalizable to  
models with non--Abelian groups, and indeed would nicely interact with  
the group theoretic content of these models, see also \cite{Singh:2009cd}. Here it will be very  
interesting to study how a non--trivial vertex or edge projector might  
change the phase structure as compared to the standard choice of the  
Haar projector. This will also facilitate a better understanding of  
the dynamics in the full gravitational models. To this end, a class of  
finite group models emulating the current gravitational EPRL models  
\cite{eprl} is constructed in \cite{bbfw}.
An important question for future research will be whether the  
degenerate phase, that is the high temperature fixed point can be  
avoided by selecting suitable projectors. This can already be studied  
for the 2D spin net models. Furthermore it has to be explored how the 
tensor network renormalization method can successfully be applied to  
three and four--dimensional spin net and spin foam models.

An alternative to using the tensor network formalism to describe the  
partition function of a system would be to apply tensor network  
renormalization as a kind of improved mean field approach in a  
canonical quantization. This would change the (statistical) systems from  $D$--dimensional classical to $(D-1)$--dimensional quantum ones. In this case the tensor networks would provide an  
ansatz for the ground states of the models \cite{Cirac,vidalprl,gulevinwen}. In  
gravity, instead of minimizing one Hamiltonian or energy functional,  
one rather has to deal with a number of constraints, which have to 
annihilate the so--called physical states. The master constraint  
\cite{Thiemann:2003zv,Dittrich:2004bn,Dittrich:2004bp} or the uniform discretization approach \cite{Campiglia:2006vy,Gambini:2009ie}  
provide a framework where just one (master) constraint has to be  
minimized.

In this work we have considered systems on a regular lattice, as this  
made explicit simulations feasible. Nevertheless one should consider   
generalizations of these methods to  random lattices \cite{Markopoulou:2000yy,Markopoulou:2002ja}.  
Another question is how the phase structure found on a fixed lattice  
relates to phases in models, that include a sum over all possible  
lattices \cite{Bonzom:2011zz,Benedetti:2011nn,Bonzom:2011ev,Carrozza:2011jn}, in particular regarding the remarks in section  
\ref{coarsemethods}.

We see this work as a contribution towards a closer link between the quantum gravity and the statistical physics / condensed matter communities. For quantum gravity researchers, this offers the prospect of new concepts and numerical tools to study the large-scale physics of their models, whereas for people working in statistical physics it offers new models, new questions, a geometrical perspective and a rich set of mathematical tools behind it.

\section*{Acknowledgements}

The authors would like to thank Benjamin Bahr for sharing with us unpublished work. We also
thank him as well as Frank Hellmann, Wojciech Kaminski and Etera Livine for enlightening
discussions. 

{\footnotesize
\bibliographystyle{utphys}
\bibliography{paper}
}

\end{document}